\documentclass[english,aps,twocolumn,showpacs]{revtex4}
\usepackage[T1]{fontenc}
\usepackage[latin9]{inputenc}
\usepackage{lipsum}
\usepackage{babel}
\usepackage{amsmath}
\usepackage{amssymb}
\usepackage{graphicx}
\usepackage{esint}
\usepackage[unicode=true,pdfusetitle,bookmarks=true,bookmarksnumbered=false,bookmarksopen=false,
 breaklinks=false,pdfborder={0 0 1},backref=false,colorlinks=false] {hyperref}

\makeatletter

\newcommand*\LyXThinSpace{\,\hspace{0pt}}
\@ifundefined{textcolor}{}
{%
 \definecolor{BLACK}{gray}{0}
 \definecolor{WHITE}{gray}{1}
 \definecolor{RED}{rgb}{1,0,0}
 \definecolor{GREEN}{rgb}{0,1,0}
 \definecolor{BLUE}{rgb}{0,0,1}
 \definecolor{CYAN}{cmyk}{1,0,0,0}
 \definecolor{MAGENTA}{cmyk}{0,1,0,0}
 \definecolor{YELLOW}{cmyk}{0,0,1,0}
}

\usepackage{babel}
\usepackage{subfigure}\usepackage{times}\usepackage{mathrsfs}\usepackage{longtable}\usepackage{latexsym}\usepackage{bm}

\newcommand{\Id}{\mathbb{I}}
\newcommand{\II}{\mathcal{I}}

\newcommand{\DD}{\mathcal{D}}

\newcommand{\jj}{\mathbf{j}}
\newcommand{\kk}{\mathbf{k}}

\newcommand{\Tr}{\mbox{Tr}}

\newcommand{\be}{\begin{equation}}
\newcommand{\ee}{\end{equation}}
\newcommand{\bee}{\begin{equation*}}
\newcommand{\eee}{\end{equation*}}
\newcommand{\ba}{\begin{align}}
\newcommand{\baa}{\begin{align*}}

\newcommand{\ket}[1]{|#1\rangle}
\newcommand{\bra}[1]{\langle #1|}

\newcommand{\average}[1]{ \langle #1 \rangle}

\newcommand{\bae}{\begin{eqnarray}}
\newcommand{\eae}{\end{eqnarray}}

\makeatother

\begin{document}

\title{Global coherence of quantum evolutions based on decoherent histories:
theory and application to photosynthetic quantum energy transport}

\author{Michele Allegra$^{1}$}

\email[]{mallegra@sissa.it}

\selectlanguage{english}%

\author{Paolo Giorda$^{2}$}

\email[]{magpaolo16@gmail.com}

\selectlanguage{english}%

\author{Seth Lloyd$^{3,4}$}

\affiliation{$^{1}$ International School for Advanced Studies (SISSA), I-34136
Trieste, Italy}

\affiliation{$^{2}$INRIM, I-10135 Torino, Italy}

\affiliation{$^{3}$Research Laboratory of Electronics, Massachusetts Institute
of Technology, Cambridge, MA 02139, USA; }

\affiliation{$^{4}$Department of Mechanical Engineering, Massachusetts Institute
of Technology, Cambridge, MA 02139, USA}
\begin{abstract}
Assessing the role of interference in natural and artificial quantum
dyanamical processes is a crucial task in quantum information theory.
To this aim, an appopriate formalism is provided by the decoherent
histories framework. While this approach has been deeply explored
from different theoretical perspectives, it still lacks of a comprehensive
set of tools able to concisely quantify the amount of coherence developed
by a given dynamics. In this paper we introduce and test different
measures of the (average) coherence present in dissipative (Markovian)
quantum evolutions, at various time scales and for different levels
of environmentally induced decoherence. In order to show the effectiveness
of the introduced tools, we apply them to a paradigmatic quantum process
where the role of coherence is being hotly debated: exciton transport
in photosynthetic complexes. To spot out the essential features that
may determine the performance of the transport we focus on a relevant
trimeric subunit of the FMO complex and we use a simplified (Haken-Strobl)
model for the system-bath interaction. Our analysis illustrates how
the high efficiency of environmentally assisted transport can be traced
back to a \textit{quantum recoil avoiding effect} on the exciton dynamics,
that preserves and sustains the benefits of the initial fast quantum
delocalization of the exciton over the network. Indeed, for intermediate
levels of decoherence, the bath is seen to selectively kill the negative
interference between different exciton pathways, while retaining the
initial positive one. The concepts and tools here developed show how
the decoherent histories approach can be used to quantify the relation
between coherence and efficiency in quantum dynamical processes.
\end{abstract}

\pacs{03.65.-w,03.65.Yz,05.60.Gg,42.50.Lc,03.67.Ac}

\maketitle

\section{Introduction}

Coherence is ultimately the most distinctive feature of quantum systems.
Finding a proper measure of the coherence present at different times
scales in a quantum dynamical system is the first essential step for
assessing the role of quantum interference in natural and artificial
processes. This is of particular relevance for those quantum evolutions
in which information or energy are transformed and transferred in
order to achieve a given task with high efficiency. In this context
the relevant questions are: how much and what kind of coherence is
created vs destroyed by the dynamical evolution? How does coherence
determine/enhance the performance of the given process? These are
in general difficult questions and to be answered they require an
appropriate and sufficiently comprehensive framework. A general and
fundamental formalism to describe quantum interference is provided
by the decoherent histories (DH) approach to quantum mechanics. DH
have mainly found applications to foundational issues of quantum mechanics
such as the formulation of a consistent framework to describe closed
quantum systems, the emergence of classical mechanics from a quantum
substrate, the solution of quantum paradoxes, decoherence theory,
quantum probabilities \cite{Griffiths,Griffiths-Book,Omnes,GMannHartle,GellMannHartle,ZurekHist,Hartle_NegProb2008,GMann-Hartle_2014-1}.
However, DH can also be a systematic tool for quantifying interference
in quantum processes, and discussing its relevance therein. Indeed,
DH provide a precise mathematical formalization of interference by
means of the the so called\emph{ decoherence matrix} $\DD$. The latter
is built on the elementary notion of \emph{histories} and allows one
to describe the quantum features vs the classical ones in tems of
interference between \emph{histories}, or \emph{pathways} if one resorts
to the mental picture of the double slit experiments. It is however
difficult to quantify in a compact and meaningful way the content
of $\DD$ and its implications for the dynamics of specific systems.
Our first main goal is therefore to define and test appropriate measures
allowing for the investigation of how interference can determine the
performance of a given quantum information processing task. Starting
from $\DD$ and by its sub-blocks we define different functionals.
In particular, we introduce a global measure of coherence $\mathcal{C}$
able to describe the coherence content of a general quantum evolution
at its various time scales; an average (over different time-scales)
measure of coherence; and average mesure of interference between
histories leading to a specific output.\par
While the tools we introduce
are of general interest and application, in order to test them we
apply them to a specific but relevant instance of quantum dynamics
taking place in photosynthetic membranes of bacteria and plants: quantum
energy transport. Here the basic common mechanism is the following:
a quantum excitation is first captured by the system and then migrates
through a network of sites (chromophores) towards a target site, e.g.,
a reaction center, where the energy is transformed and used to trigger
further chemical reactions. There is now an emerging consensus that
efficient transport in natural and biologically-inspired artificial
light-harvesting systems builds on a finely tuned balance of quantum
coherence and decoherence caused by environmental noise \cite{Ishizaki-PCCP,Ishizaki-PNAS,Belcher-Virus,Kevin-Bardeen_Original,GiordaJCP},
a phenomenon known as environment-assisted quantum transport (ENAQT).
This paradigm has emerged with clarity in recent years, as modern spectroscopic
techniques first suggested that exciton transport within photosynthetic
complexes might be coherent over appreciable timescales~\cite{Engel}.
Indeed, a growing number of experiments has provided solid evidence
that coherent dynamics occurs even at room temperature for unusually
long timescales (of the order of $100fs$)~\cite{Collini,Panit}.
Efforts to describe these systems have led to general models of ENAQT~\cite{Kevin-Bardeen_Original,Mohseni,Mohseni2,Plenio,Cao},
depicting the complex interplay of three key factors: coherent motion,
i.e., quantum delocalization of the excitation over different sites,
environmental decoherence, and localization caused by a disordered
energy landscape. So far, the presence of coherence in light-harvesting
systems has been qualitatively associated to the observation of distinctive
`quantum features'. Originally, coherence was identified with `quantum
wavelike' behavior as reflected by quantum beats in the dynamics of
chromophore populations within a photosynthetic complex. Later works,
employing quantum-information concepts and techniques, have switched
attention towards quantum correlations between chromophores, in particular
quantum entanglement\cite{Caruso,Whaley,GiordaLHCII,Ishizaki}. Besides
being open to criticism (see, e.g.,~\cite{Miller,Tiersch}), these
approaches do not provide direct quantitative measures of coherence
in the presence of noise. Therefore, in what follows, we shall apply
the novel tools based on DH to a simple yet fundamental model of quantum
energy transfer. We will focus on a relevant trimeric subunit of the
Fenna-Matthews-Olson (FMO) complex, the first pigment-protein complex
to be structurally characterized \cite{Blankenship}. The trimer is
virtually the simplest paradigmatic model retaining the basic charcteristics
of a disordered transfer network and it can also be conceived as an
essential building block of larger networks. For simplicity, we will
use the well-known Haken-Strobl model~\cite{Haken} to describe the
interplay between Hamiltonian and dephasing dynamics. While the model
is an oversimplified description of the actual dynamics taking place
in real systems, it allows to spot out the essential features that
may determine the high efficiency of the transport. We shall initially
focus on a new coherence measure $\mathcal{C}$, based on the decoherence
matrix, and characterize its behavior verifying that it can consistently
identify the bases and timescales over which quantum coherent phenomena
are present during the evolution of the system. We shall then show how the
average coherence exhibited on those time scales can be connected
with the delocalization process. A more detailed analysis will be
aimed at distinguishing between constructive and destructive interference
affecting the histories ending at the site where the excitation exits
the photosynthetic structure. By using the decoherence functional,
we will show that the beneficial role of dephasing for the transport
efficiency lies in a selective suppression of destructive interference,
a fact that has been systematically suggested in the literature, but
never expressed within a general and comprehensive framework that
allows the quantitative evaluation of coherence and its effects.\par
The application of the introduced tools and methods based on DH to
a simple yet paradigmatic system shows how one can properly quantify
the coherence content of a complex quantum dynamics and elucidate the
role of coherence in determining the overall efficiency of the process~\cite{Pleniocoherence}.\par
The paper is structured as follows. In Section \ref{sec:Decoherent-histories}
we review the basic decoherent histories formalism. In Section \ref{sec:The-coherence-measure}
we define the measure of coherence $\mathcal{C}$ and describe its
meaning and properties. In Section \ref{sec:Trimer} we first introduce
the used model for describing the energy transport in the selected
trimeric complex. We then discuss the coherence properties of the
excitonic transport: By means of the appropriate measures based on
the decoherent histories formalism we identify the essential features
that may determine the high efficiency of the transport. In Section
\ref{sec:FMO} we briefly discuss how to extend our results to the
whole FMO complex. In Section \ref{sec:Conclusions} we summarize
our results an draw our conclusions.

\section{Decoherent histories\label{sec:Decoherent-histories}}

The formalism of decoherent (or consistent) histories was developed
in slightly different flavors by Griffiths~\cite{Griffiths,Griffiths-Book},
Gell-Mann~\cite{GMannHartle,GMann-Hartle_2014-1}, Hartle\cite{Hartle_NegProb2008}
and Omn\`es~\cite{Omnes}. DH provide a consistent formulation of quantum
mechanics where %with no reference to measurements, %measurements play no fundamental role,
%which allows to apply the theory to closed systems -- in particular, to the whole universe
%(quantum cosmology). 
%In the DH framework, the physics is essentially the same as in 
%the standard formalism, but 
%experiments are replaced by the consistency condition and 
probabilities of measurement outcomes are replaced by probabilities
of \textit{histories}. In this formulation, external measurement apparatuses
are not needed, and then one does not need to postulate a ``classical
domain\textquotedbl{} of observers. As a consequence, quantum mechanics
becomes a theory that allows the calculation of probabilities of sequences
of events within any closed system, including the whole universe,
without the necessity of invoking postulates about the role of measurement.
In this framework, the ``classical domain\textquotedbl{} can be seen
to emerge as the description of the system becomes more and more coarse-grained.\par
 The idea of `histories' stems from Feynman's `sum-over-histories'
formulation of quantum mechanics. As is known, any amplitude $\langle\psi_{f}|U(t_{f}-t_{0})|\psi_{i}\rangle$
between an initial and a final state can be expressed as a sum over
paths, or histories: Upon inserting the identity decomposition $\mathbb{I}=\sum_{j}|j\rangle\langle j|=\sum_{j}P_{j}$
at differerent times $t_{1}\dots t_{N}$ we get 
\begin{small}
\begin{align*}
 \langle\psi_{f}|U(t_{f}-t_{0})|\psi_{i}\rangle= & \langle\psi_{f}|U(t_{f}-t_{N})\sum_{j_{N}}P_{j_{N}} U(t_{N}-t_{N-1}) \dots \\
\ & \dots U(t_{2}-t_{1})\sum_{j_{1}}P_{j_{1}}U(t_{1}-t_{0})|\psi_{i}\rangle= \\
= \sum_{j_{1\dots}j_{N}}\langle\psi_{f}|P_{j_{N}}(t_{N}) & \dots P_{j_{1}}(t_{1})|\psi_{i}\rangle %\quad ,
\end{align*}
\end{small}
where we use the Heisenberg notation $P_{j}(t)=U^{\dagger}(t-t_{0})P_{j}U(t-t_{0})$.
Thus the total amplitude $\langle\psi_{f}|U(t_{f}-t_{0})|\psi_{i}\rangle$
is decomposed as a sum of amplitudes, each one corresponding to a
different \textit{history} identified by a sequence of projectors
$P_{j_{N}}\dots P_{j_{1}}$. \par
 The decoherent histories formalism assumes that histories are the
fundamental objects of quantum theory and gives a prescription to
attribute probabilities to (sets of) histories. A history is defined
as a sequence of projectors %$P^{1},\dots,P^{N}$
at times $t_{1}<\dots<	t_{N}$. %Predictions
%are formulated in terms of
Probabilites can be assigned within exhaustive sets of exclusive histories,
i.e., sets of histories $\mathcal{S}_{N}=\{t_{1},\dots,t_{N},P_{j_{1}},\dots,P_{j_{N}}\}$
where subscripts $j_{1},\dots,j_{N}$ label different alternatives
at times $t_{1},\dots,t_{N}$. Histories are exhaustive and exclusive
in the sense that the projectors at each time %$t_{\ell}$ 
satisfy relations of orthogonality $P_{j}P_{k}=\delta_{jk}P_{j}$,
and completeness, $\sum_{j}P_{j}=\mathbb{I}$. % relations.
In other words, the projectors $P_{j}$ define a %partition of the identity,or a 
projective measurement. Within a specified set, any history can be
identified with the sequence of alternatives $\jj\equiv j_{1},\dots,j_{N}$
realized at times $t_{1},\dots,t_{N}$. \par
 %If the $P_{j}$ are one-dimensional, histories are called \emph{fine-grained}. %else they are \emph{coarse-grained}.
%A less detailed description  can be obtained through a procedure called coarse-graining, whereby 
Different alternative histories can be grouped together with a procedure
called \textit{coarse-graining}. Starting from histories $\jj$ and
$\kk$ we can define a new, \textit{coarse-grained} history $\mathbf{m}=\mathbf{j}\lor\mathbf{k}$
by summing projectors for all times $t_{\ell}$ such that $j_{\ell}$
and $k_{\ell}$ differ: 
\begin{align*}
P_{m_{\ell}}=P_{j_{\ell}}+P_{k_{\ell}} & \mbox{ if }j_{\ell}\neq k_{\ell}\\
P_{m_{\ell}}=P_{j_{\ell}} & \mbox{ if }j_{\ell}=k_{\ell},
\end{align*}
for all $\ell=1,\dots,N$. By iterating this procedure, one can obtain
more and more coarse-grained histories. %Analogously, 
%starting from $n$ histories $\jj_1 \dots \jj_n $ we can define a new history $\mathbf{m}=\lor_{} \mathbf{j_}\lor\mathbf{k}$  
%by summing projectors for all times, $P_{m_{\ell}}=P_{j_{\ell}}+P_{k_{\ell}}$.
A special type of coarse-graining is the \textit{temporal coarse-graining}:
we group together histories $\jj,\kk,\dots,\mathbf{l}$ such that
such that at some time $t_{\ell}$ we have $P_{j_{\ell}}+ P_{k_{\ell}}+\dots P_{l_{\ell}}=\Id$.
Then the coarse-grained history $\mathbf{m}=\jj\lor\kk\lor\dots\lor\mathbf{l}$
contains only one-projector (equal to the identity) at time $t_{\ell}$,
that can be neglected and hence removed from the string of projectors
defining the history. On the other hand \textit{temporal fine-graining}
can be implemented for example by allowing different alternatives
at a times $t_{k}\notin\{t_{1},..,t_{N}\}$. In particular, one can
create new sets of histories $\mathcal{S}_{N+1}=\{t_{1},\dots,t_{N},t_{N+1},P_{j_{1}},\dots,P_{j_{N}},P_{j_{N+1}}\}$
from a given one $\mathcal{S}_{N}$ by adding different alternatives
at time $t_{N+1}>t_{N}$; the sets $\mathcal{S}_{N+1}$ are fine grained versions
of the sets $\mathcal{S}_{N}$. \\
 %To introduce probabilities, we must
Once we specify the initial state $\varrho$ and the (unitary) time
evolution $U(t)$, %\equiv j_1,\dots, j_N$ (we identify histories with the sequence of alternatives
%$j_\ell$ realized at times $t_\ell$)
we can assign any history $\jj$ a \textit{weight} 
\[
w_{\jj}=\Tr[C_{\jj}\varrho C_{\jj}^{\dagger}],\quad\mbox{with}\quad C_{\jj}=P_{j_{N}}(t_{N})\dots P_{j_{1}}(t_{1})
\]
where we use the Heisenberg notation $P_{j_{\ell}}(t_{\ell})=U(t_{\ell})P_{j_{\ell}}U(t_{\ell})^{\dagger}$.
When the initial state is pure, $\varrho=\ket{\psi_{i}}\bra{\psi_{i}}$
and the final projectors are one dimensional, $P_{j_{N}}=\ket{\psi_{j_{N}}}\bra{\psi_{j_{N}}}$,
this formula takes the simple form of a squared amplitude 
\begin{equation}
w_{\jj}=|\bra{\psi_{j_{N}}}P_{j_{N-1}}(t_{N-1})\dots P_{j_{1}}(t_{1})\ket{\psi_{i}}|^{2}.
\end{equation}
Weights cannot be interpreted as true probabilities, in general. %only if the decoherence condition is satisfied. 
Indeed, due to quantum interference between histories, the $w_{j}$
do not behave as classical probabilities. Indeed, consider two exclusive
histories $\jj,\kk\in\mathcal{S}$ and the relative coarse-grained
history $\mathbf{m}=\jj\lor\kk$ by: $P_{m_{\ell}}=P_{j_{\ell}}+P_{k_{\ell}},\forall\ell$.
If the $w_{\jj}$ were real probabilities, we would expect $w_{\mathbf{m}}=w_{\jj}+w_{\kk}$.
Instead, what we find is 
\[
w_{\mathbf{m}}=w_{\jj}+w_{\kk}+2Re(\Tr[C_{\jj}\varrho C_{\kk}^{\dagger}]).	
\]
Due to the non-classical term $Re(\Tr[C_{\jj}\varrho C_{\kk}^{\dagger}])$,
representing quantum interference between the histories $\jj$ and
$\kk$, the classical probability-sum-rule is violated. %Only if $ Re (\Tr[C_{\jj}\varrho C_{\kk}^{\dagger}])=0$ can we interpret $p_\jj$ and $p_\kk$ as probabilities.
The matrix 
%\lipsum[1]
\begin{widetext}
\begin{equation}
\DD_{\jj\kk}=\Tr[C_{\jj}\varrho C_{\kk}^{\dagger}]=\Tr[P_{j_{N}}U(t_{N}-t_{N-1})\dots P_{j_{1}}U(t_{1})\varrho U(t_{1})^{\dagger}P_{k_{1}}\dots U(t_{N}-t_{N-1})^{\dagger}P_{k_{N}}]\label{Eq: decomatrix}
\end{equation}
\end{widetext}
%\lipsum[1]
is called \emph{decoherence functional} or \emph{decoherence matrix}.
The decoherence matrix can be thought of as a ``density
matrix over histories'': Its diagonal elements are the
weights of histories and its off-diagonal elements are interferences
between pairs of histories. The decoherence matrix has the following
properties: i) it is Hermitian ii) it is semipositive definite iii)
it is trace one iv) it is block-diagonal in the last index, $\DD_{\jj\kk}=\delta_{j_{N}k_{N}}\DD_{\jj\kk}$.
Weights of coarse-grained histories can be obtained by summing matrix
entries in an $n\times n$ block of the decoherence matrix corresponding
to the original fine-grained histories. For istance, the weight of
history $\mathbf{m}=\jj\lor\kk$ is obtained by summing entries of
a $2\times2$ block of the decoherence matrix: 
\[
w_{\mathbf{m}}=\DD_{\jj\jj}+\DD_{\kk\kk}+\DD_{\jj\kk}+\DD_{\kk\jj}.
\]
A necessary and sufficient condition to guarantee that the probability
sum rule $w_{\jj\lor\kk}=w_{\jj}+w_{\kk}$ apply within a set of histories
is

\[
Re[\DD_{\jj\kk}]=0,\forall\jj\neq\kk.
\]
This condition is termed as \emph{weak decoherence\cite{GellMannHartle}};
the necessary and sufficient condition that is typically satisfied
\emph{\cite{GellMannHartle}} and that we will adopt in the following
is the stronger one termed as \emph{medium decoherence} 
\begin{equation}
\DD_{\jj\kk}=0,\forall\jj\neq\kk.\label{Eq.: mediumdeco}
\end{equation}
%\emph{weak decoherence} $Re(\DD_{\jj \kk})=0,\forall \jj \neq \kk$
Medium decoherence implies weak decoherence. Any exhaustive and set
of exclusive histories satisfying medium decoherence is called a \textit{decoherent
set}. The fundamental rule of DH approach is that probabilities can
be assigned within a decoherent set, each history being assigned a
probabability equal to its weight. If medium decoherence holds, the
diagonal elements of the decoherence matrix can be identified as real
probabilities for histories and we can write $\DD_{\jj\jj}=p_{\jj}$.
\\
 \\
 %Upon coarse-graining, we obtain ...
Due to property iv), if we perform a temporal coarse-graining over
all times except the last, we obtain `histories' with only one projection,
$P_{j_{N}}$ at the final time $t_{N}$. These histories automatically
satisfy medium decoherence: 
\begin{align*}
\sum_{j_{1},\dots,j_{N-1}}\sum_{k_{1},\dots k_{N-1}}\DD_{\jj\kk} & = \\
 = \delta_{j_{N}k_{N}}\Tr[P_{j_{N}}(t_{N})\varrho P_{j_{N}}(t_{N})] &  \equiv\delta_{j_{N}k_{N}}p_{j_{N}}
\end{align*}
where $p_{j_{N}}\equiv\Tr[P_{j_{N}}(t_{N})\varrho P_{j_{N}}(t_{N})]$
is the probability that the system is in $j_{N}$ at time $t_{N}$.
Due to interference, the probability of being in $j_{N}$ at time
$t_{N}$ is \textit{not} simply the sum of probabilities of all alternative
paths leading to $j_{N}$, i.e, of all alternative histories with
final projection $P_{j_{N}}$. In formulas, % cannot be obtained 
%by summing the weights of all histories leading to $j_N$ at time $t_N$:
\[
p_{j_{N}}\neq\sum_{j_{1},\dots j_{N-1}}w_{\jj}=\sum_{j_{1},\dots j_{N-1}}\DD_{\jj\jj}.
\]
%Upon defining the sum of w w_{j_N} \equiv $
The probability and the global interference of histories $\mathcal{I}_{j_{N}}(\tau)$
ending in $j_{N}$ can be thus expressed as 
\begin{equation}
p_{j_{N}}(\tau)=\sum_{j_{1},\dots j_{N-1}}w_{\jj}(\tau)+\mathcal{I}_{j_{N}}(\tau)\label{eq: interference and probability at ending site j}
\end{equation}
with $\tau=N\Delta t\,$. Destructive interference will happen when
$\mathcal{I}_{j_{N}}<0$, constructive interference when $\mathcal{I}_{j_{N}}>0$.
\par
 %$\bar{P}^{\ell}_1=\sum_{j_\ell \in \Sigma^\ell_1 }P_{j_\ell}^{\ell}=\mathbb{I}$.
%% that is the sum of all fine-grained projectors and sums up to the identity.
%%Then histories in $\Sigma'$ contain no event at time $t_{\ell}$, because $\mathbb{I}$ is a trivial event
%that can be neglected and hence removed from the string of projectors defining the history.
%(some Authors consider as fundamental the weak decoherence condition $\mbox{Re}(\Tr[\DD_{\jj\kk}])=0$ which is
%necessary and sufficient to ensure the probability-sum-rule; apart from the fact that in most physical cases weak 
%decoherence only comes together with medium decoherence~\cite{GMannHartle}, weak decoherence has many shortcomings 
%-- for a review, see~\cite{AllegraTesi}). \\
%Usually medium decoherence (approximately) occurs as a result of coarse-graining. Among the types of coarse-graining that can 
%lead to medium decoherence, a common one is the type that involves 
The decoherent histories formalism is consistent with and encompasses
the model of environmentally induced decoherence\cite{ZurekHist}.
Given a factorization of the Hilbert space into a subsystem of interest
and the rest (environment), $\mathcal{H}=\mathcal{H}_{S}\otimes\mathcal{H}_{\mathcal{E}}$,
the events of a history %$\jj$ 
take the form $P_{j_{\ell}{j'}_{\ell}}=\tilde{P}_{j_{\ell}}\otimes\Pi_{{j'}_{\ell}}$
where $\tilde{P}_{j_{\ell}}$ and $\Pi_{j'_{\ell}}$ are projectors
onto Hilbert subspaces of $\mathcal{H}_{S}$ and $\mathcal{H}_{\mathcal{E}}$
respectively. Histories for $S$ alone can be obtained upon considering
appropriate coarse-grainings over the degrees of freedom of the environment,
such that the events are $\tilde{P}_{j_{\ell}}\otimes\Id_{\mathcal{E}}$
where $\Id_{\mathcal{E}}$ is the identity over $\mathcal{H}_{\mathcal{E}}$.
%Given a factorization $\mathcal{H}=\mathcal{H}_{S}\otimes\mathcal{H}_{\mathcal{E}}$, and considering
%a coarse graining over the environment degrees of freedom, such that
%the events are $\tilde{P}_{j_\ell}^{\ell}\otimes\Id_{\mathcal{E}}$, the decoherence matrix for $S$ reads:
%\bee
%\DD_{\jj \kk} = \Tr[\tilde{P}_{j_N}U(t_N-t_{N-1})\dots \tilde{P}_{j_1} U(t_1-t_0) \varrho(t_0) U(t_1-t_0)^\dagger \tilde{P}_{k_1} \dots U(t_N-t_{N-1}) \tilde{P}_{k_N}]
%\eee
%where $U$ is the joint evolution of system and environment.
%For convenience, %let us first rewrite the decoherence matrix in the Schrodinger picture, assigning the time dependence to the density matrixlet us first introduce 
Upon introducing the time-evolution propagator $\mathcal{K}_{t\ t_{0}}$
as $\varrho(t)=U(t-t_{0})\varrho(t_{0})U(t-t_{0})^{\dagger}\equiv\mathcal{K}_{t\,t_{0}}[\varrho(t_{0})]$,
we can rewrite the decoherence matrix as: 
\begin{align}
\mathcal{D}_{\jj\kk}= & \Tr[\tilde{P}_{j_{N}}\mathcal{K}_{t_{N}\ t_{N-1}}[\tilde{P}_{j_{N-1}}\mathcal{K}_{t_{N-1}\ t_{N-2}}[\label{Eq: dmschrodinger}\\
 & \dots\mathcal{K}_{t_{1}\ t_{0}}[\varrho_{0}]\dots]\tilde{P}_{k_{N-1}}]\tilde{P}_{k_{N}}]. \nonumber 
\end{align}
If the initial state is factorized $\varrho(t_{0})=\tilde{\varrho}_{S}(t_{0})\otimes\varrho_{\mathcal{E}}(t_{0})$,
then the reduced density matrix $\varrho_{S}(t)=Tr_{\mathcal{E}}[\varrho(t)]$
evolves according to $\varrho_{S}(t)=\tilde{\mathcal{K}}_{t\ t_{0}}\varrho_{S}(t_{0})$
where $\tilde{\mathcal{K}}$ is the (non-unitary) reduced propagator\index{reduced propagator}
defined by
\begin{equation}
\begin{small}
\Tr_{\mathcal{E}}[U(t-t_{0})\tilde{\varrho_{S}}(t_{0})\otimes\varrho_{\mathcal{E}}(t_{0})U^{\dagger}(t-t_{0})]=\tilde{\mathcal{K}}_{t\ t_{0}}[\tilde{\varrho}_S(t_{0})].
\end{small}
\end{equation}
If the evolution of the system and environment is Markovian, %the semigroup property holds $ \tilde{K}_{t_2 \ t_0 } = \tilde{\mathcal{K}}_{t_2 \ t_1 } \tilde{\mathcal{K}}_{t_1 \ t_0 } $ and 
we can write $\tilde{\mathcal{K}}_{t\ t'}=\tilde{\mathcal{K}}_{t-t'}$.
As proved by Zurek~\cite{ZurekHist}, under the assumption of Markovianity
we can rewrite the decoherence matrix in terms of reduced quantities
alone, i.e., quantities pertaining to the system only: 
\begin{widetext}
\begin{align}
\mathcal{D}_{\jj\kk}= & \Tr_{S}[\tilde{P}_{j_{N}}\tilde{\mathcal{K}}_{t_{N}\ t_{N-1}}[\tilde{P}_{j_{N-1}}\tilde{\mathcal{K}}_{t_{N-1}\ t_{N-2}}[\dots\tilde{\mathcal{K}}_{t_{1}\ t_{0}}[\varrho_{0}]\dots]\tilde{P}_{k_{N-1}}]\tilde{P}_{k_{N}}]. \label{Eq: reduceddec}
\end{align}
\end{widetext}
%Obviously, the system evolution can be Markovian only over time intervals $\Delta t \gg \tau_C$ where $\tau_C$ is the 
%system-environment correlation time\index{correlation time}.
%Thus, Eq. (\ref{Eq: reduceddec}) is valid only if the time interval $\Delta t$ between projections 
%(the temporal coarse-graining) is much higher than $t_C$. Notice that usually $t_C \ll t_D$.
%\\
%Even though generic histories for the (closed) compound system 
%would form a coherent set, coarse-grained histories for the system alone can decohere.
%In several cases, depending on the dynamics, the coarse-graining over
%the environment yields indeed a decoherent set of histories for the
%system for an appropriate choice of the $\tilde{P}_{j_\ell}^{\ell}$. In
%particular, in many physical cases the resulting decoherent histories
%are composed by projectors in some given, \emph{fixed} basis. 
That is, the model of environmentally induced decoherence can be obtained
by applying the decoherent histories formalism to system and environment
together, and by coarse-graining over the degrees of freedom of the
environment.

\section{The coherence measure $C$\label{sec:The-coherence-measure}}

The DH approach provides the most fundamental framework in which the
transition from the quantum to the classical realm can be expressed. Indeed,
it is based on the most basic feature characterizing the quantum
world: interference and the resulting coherence of the dynamical evolution.
Despite being a well developed field of study, the DH history approach
lacks for a proper global measure of the coherence produced by the
dynamics at the different time scales. We therefore introduce a measure
that quantifies the global amount of coherence within a set of histories.
Assume projectors for all times $t_{\ell},\ell=1,\dots,N$ are taken
in a fixed basis % histories with projections in a fixed basis.To any basis we can associate the projectors
$\ket{e_{j}}$, $P_{j}^{\ell}=\ket{e_{j}}\bra{e_{j}}$. %To any given projective measurement $P_j$ ($P_j P_k = \delta_{jk}$, $\sum_j P_j = \mathbb{I}$) we can naturally associate an exhaustive set of exclusive
%histories. Consider then a set of 
Assume further that histories are composed by taking equally spaced
times between consecutive projections i.e., $t_{1}=\Delta t,\dots,t_{N}=N\Delta t$.
% and alternatives for all times specified bythe the same projectors $P_j$:  $P^\ell_{j_\ell}=P_{j_\ell}$. 
(in other words, histories correspond to projections applied in
the same basis and repeated at regular times). For such a set of histories,
consider the decoherence matrix 
\[
\DD_{\jj\kk}^{(N,P,\Delta t)}=\Tr[C_{\jj}^{(N,P,\Delta t)}\varrho C_{\kk}^{(N,P,\Delta t)}]
\]
where $C_{\jj}^{(N,P,\Delta t)}=P_{j_{N}}(N\Delta t)\dots P_{j_{1}}(\Delta t)$.
%We can now define a partition-dependent dynamical entropy as the von Neumann entropy
Take the von Neumann entropy of the decoherence matrix, 
\begin{align}
h(P,N,\Delta t)=-\Tr[\DD^{(N,P,\Delta t)}\log\DD^{(N,P,\Delta t)}].
\end{align}
%representing the amount of information produced by the repeated $P$ measurements, 
%we can define a partition-dependent dynamical entropy as $\frac{1}{N}h_N(P)$. Notice that $h_N$ is a measure of quantum information since $\DD^{(N,P)}$
%includes coherent correlations between measurement results. 
%Upon maximising the (asymptotic) information production rate over projective measurements~\cite{Sethgeneralized},
%we retrieve the ALF entropy 
%\begin{equation}
%h_{ALF}=\frac{1}{N} \lim_{N\to\infty}\max_{P} h_{N}(P) 
%\end{equation}
%that was defined in~\cite{Alicki1}. %by translating the KS construction in the quantum language.
Due to coherence between histories, $h_{N}$ differs from the `classical-like'
Shannon entropy of history weights %of measurement results 
\begin{equation}
h^{(c)}(P,N,\Delta t)=-\sum_{\jj}w_{\jj}^{(N,P,\Delta t)}\log w_{\jj}^{(N,P,\Delta t)}
\end{equation}
where $w_{\jj}^{(N,P,\Delta t)}=\Tr[C_{\jj}^{(N,P,\Delta t)}\varrho C_{\jj}^{(N,P,\Delta t)}]$
are the diagonal elements of $\DD^{(N,P,\Delta t)}$, i.e., the weights.
%However, the stronger %the set of 
%the decoherence (the smaller the off-diagonal elements of $\DD^{(N,P)}$), the more the two quantities approach each other. 
%In fact, the quantity 
The difference between the two quantities is wider if off-diagonal
elements of the decoherence matrix are bigger, i.e., if the set of
histories is more coherent. Let us define:

\begin{equation}
C(P,N,\Delta t)\equiv\frac{h^{(c)}(P,N,\Delta t)-h(P,N,\Delta t)}{h^{(c)}(P,N,\Delta t)}. \label{eq: Cfunction}
\end{equation}
We argue that $C(P,N,\Delta t)$ is suitable to be used as a general
measure of coherence within the set of histories defined by $P,N,\Delta t$.
Indeed, we can readily prove the following properties:  \\
i) $0\leq C(P,N,\Delta t)<1$. $C(P,N,\Delta t)<1$ is obvious. To prove $C(P,N,\Delta t)>0$, let
us define a matrix $\tilde{\mathcal{D}}_{\jj\kk}^{(N,P,\Delta t)}=\delta_{\jj\kk}\mathcal{D}_{\jj\kk}^{(N,P,\Delta t)}$
where off-diagonal entries are set to zero. Since 
\begin{align*}
\Tr[\mathcal{D}^{(N)}\log\tilde{\mathcal{D}}^{(N)}] & =\sum_{\jj}D_{\jj,\jj}^{(N,P,\Delta t)}\log D_{\jj,\jj}^{(N,P,\Delta t)}= \\
 & = \Tr[\tilde{\mathcal{D}}^{(N,P,\Delta t)}\log\tilde{\mathcal{D}}^{(N,P,\Delta t)}]
\end{align*}
we obtain that the numerator of (\ref{eq: Cfunction}) can be expressed
as a quantum relative entropy: 
\begin{align*}
 & \ h^{(c)}(P,N,\Delta t)-h(P,N,\Delta t)=\\
 & -\Tr[\tilde{\DD}^{(N,P,\Delta t)}\log\tilde{\DD}^{(N,P,\Delta t)}]+\Tr[\DD^{(N,P,\Delta t)}\log\DD^{(N,P,\Delta t)}] \\
& =\Tr[\tilde{\mathcal{D}}^{(N,P,\Delta t)}(\log\mathcal{D}^{(N,P,\Delta t)}-\log\tilde{\mathcal{D}}^{(N,P,\Delta t)})]\\
 & =h(\mathcal{D}^{(N,P,\Delta t)}||\tilde{\mathcal{D}}^{(N,P,\Delta t)})\geq0
\end{align*}
where $h(A||B)\geq0$ is the relative entropy between $A$ and $B$. \\
ii) $C(P,N,\Delta t)=0$ iff $\mathcal{D}_{\jj,\jj}^{(N,P,\Delta t)}=\tilde{\mathcal{D}}_{\jj,\jj}^{(N,P,\Delta t)}$,
i.e., $C(P,N,\Delta t)$ vanishes if medium decoherence holds for
the set of histories, since the two quantities $h(N,P,\Delta t)$
and $h^{(c)}(N,P,\Delta t)$ coincide in this case. \par
Thus $C(P,N,\Delta t)$ is in essence a (statistical) distance between
the decoherence matrix $\DD$ and the corresponding diagonal matrix
$\tilde{\DD}^{(P,N,\Delta t)}$, renormalized so that its value lies
between $0$ and $1$. The greater are the off-diagonal elements of
$\DD^{(P,N,\Delta t)}$, the greater the distance. The meaning of
$C(P,N,\Delta t)$ can be easily understood if we use the linear entropy, 
a lower bound to the logarithmic version: 
\begin{align*}
\ & h_{L}(P,N,\Delta t)=1-\Tr[(\DD^{(N,P,\Delta t)})^{2}],\\
\ & 1-h_{L}^{(c)}(P,N,\Delta t)=\Tr[(\tilde{\DD}^{(N,P,\Delta t)})^{2}].
\end{align*}
In this case, we obtain a `linear entropy' proxy of $C(N,P,\Delta t)$
as: 
\begin{align*}
& C_{L}(P,N,\Delta t)\equiv\frac{h_{L}^{(c)}(P,N,\Delta t)-h_{L}(P,N,\Delta t)}{h_{L}^{(c)}(P,N,\Delta t)}=\\
& = \frac{\sum_{\jj\neq\kk}|\DD_{\jj\kk}^{(N,P,\Delta t)})|^{2}}{1-\sum_{\jj}|\DD_{\jj\jj}^{(N,P,\Delta t)})|^{2}}
\end{align*}
which is a simplified version that, by avoiding the diagonalization
of $\DD^{(N,P,\Delta t)}$, helps containing the numerical complexity.

The measure introduced is well grounded on physical considerations.
In the following we will apply it to a simple system in order to check
its consistency, and later use it to characterize the coherence
properties of the evolutions induced by various regimes of interaction
with the environment. First, one has to check whether the measure properly
takes into account the action of the bath. In particular, if the bath
is characterized by a decoherence time $\gamma^{-1}$, it is known
(\cite{ZurekHist}) that on time scales $\Delta t\ge\gamma^{-1}$
the decoherence matrix becomes diagonal: The probability of a history
at time $t_{N+1}$ can be fully determined by its probability at time
$t_{N}$, since no interference can occur between different histories.
Indeed, the action of the bath is to create a\emph{ decoherent set
of histories} that are defined by a proper projection basis: the pointer
basis (\cite{ZurekHist}). Therefore, the fine-graining procedure obtained
by constructing a set of histories $\mathcal{S}_{N+1}$ via the addition
of a new complete set of projections in the same basis at time $t_{N+1}=(N+1)\Delta t$
to the set $\mathcal{S}_{N}$, should leave the coherence functional
$C$ invariant, i.e., $C(P,N+1,\Delta t)\approx C(P,N,\Delta t)$.
If instead $\Delta t<\gamma^{-1}$ the same fine-graining procedure
should lead to $C(P,N+1,\Delta t)\ge C(P,N,\Delta t)$.

Before passing to analyze a specific system, we want to focus on
the complexity of the evaluation of $\DD$ and $\mathcal{{C}}$.
The dimension of the decoherent matrix grows with the dimension $d$
of the basis $P$ and the number $N$ of time instants that define
each history as $d^{2N}$. This exponential growth in principle limits
the application of the DH approach to small systems. However, as for
the system considered in this paper the computational effort is contained
due to the small number of subsystems (chromophores) and the small
dimension of the Hilbert space which is limited to the single-exciton
manifold. As we shall see, by limiting the choice of $N$ to a reasonable
number, the analysis can be fruitfully carried even on a laptop.

\section{Trimer\label{sec:Trimer}}

We now start to analyze decoherent histories in simple models of energy
transfer %small networks of $d$ chromophores (sites).
comprising a small number $d$ of chromophores (sites). Neglecting
higher excitations, each site $i$ can be in its ground $\ket{0}_{i}$
or excited $\ket{1}_{i}$ state. We work in the single-excitation
manifold, and define the site basis as 
\[
\ket{i}\equiv\ket{0}_{1}\dots\ket{1}_{i}\dots\ket{0}_{d}\qquad i=1\dots d,
\]
i.e., state $\ket{i}$ represents the exciton localized at site $i$.
On-site energies and couplings are represented by a Hamiltonian $H$
that is responsible for the unitary part of the dynamics. Interaction
with the environment is implemented by the Haken-Strobl model, that
has been extensively used in models of ENAQT~\cite{Mohseni,Plenio,GiordaLHCII,Cao}.
The effect of the environment is represented by a Markovian dephasing
in the site basis, expressed by Lindblad terms $L$ in the evolution,
as follows: 
\begin{equation}
\dot{\varrho}=[H,\varrho]+\sum_{i}\gamma_{i}[2L_{i}\varrho L_{i}^{\dag}-L_{i}^{\dag}L_{i}\varrho-\varrho L_{i}^{\dag}L_{i}]\label{eq: HakenStrobl}
\end{equation}
where $L_{i}=|i\rangle\langle i|$ are projectors onto the site basis,
and $\gamma_{i}$ are the (local) dephasing rates. Furthermore, site
$d$ can be incoherently coupled to an exciton sink, represented by
a Linblad term
\[
k_{trap}[2L_{trap}\varrho L_{trap}^{\dag}-L_{trap}^{\dag}L_{trap}\varrho-\varrho L_{trap}^{\dag}L_{trap}]
\]
where $L_{trap}=|sink\rangle\langle e|$ and $k_{trap}$ is the trapping
rate. Contrary to other works, we neglect exciton recombination, as
it acts on much longer timescales $(\sim1\:ns)$ than dephasing and trapping.
\\
 The global evolution is Markovian and can be represented by means
of the Liouville equation

\begin{equation}
\dot{\varrho}=\mathcal{L}(\varrho)=\mathcal{L}_{H}(\varrho)+\mathcal{L}_{\gamma}(\varrho)+\mathcal{L}_{{trap}}(\varrho)
\end{equation}
that can be simply solved by exponentiation,

\begin{equation}
\varrho(t)=e^{i\mathcal{L}t}(\varrho(0)).
\end{equation}
In the notation above, the propagator has the form $\tilde{\mathcal{K}}_{t't}=e^{i\mathcal{L}(t'-t)}$.
The efficiency of the transport can be evaluated as the leak of the
population $p_{e}(t)=\bra{e}\rho\ket{e}$ of the exit site $e$ towards
the sink: 
\begin{equation}
\eta(t)=2k_{trap}\intop_{0}^{t}\bra{e}\rho\ket{e}.\label{eq: efficiency}
\end{equation}
The overall efficiency of the process is obtained by letting $t\rightarrow\infty.$

While its Markovianity limits the faithful description
of decoherence processes actually taking place in real photosynthetic
systems,\emph{ }the model retains the basic and commonly accepted
aspects of decoherence, that acts in the site basis: albeit in a complex
non-Markovian way, the protein enviroment measures the system locally
(i.e., on each site), thus\emph{ }destroying the coherence in the site basis
and creating it in the exciton basis. Note that the formalism can also
be applied to a `dressed' or polaronic basis where we include strong
interactions between chromophores and vibrational modes. That is,
to apply the DH method, one only needs a model in which an exciton
hops between sites, dressed or undressed. The model is therefore suitable
to readily implement the decoherent histories paradigm and to spot
the main basic features we are interested in and that are at the basis
of the success of ENAQT.

The FMO unit has $7$ chromophores and a complex energy and coupling
landscape with no symmetries. Energies and couplings (i.e., the Hamiltonian
$H$) can be obtained by different techniques: They can be extracted
by means of 2D spectroscopy as in~\cite{Cho} or computed through
ab initio calculations as in~\cite{Renger}, with similar but not
exactly equal results. This very complex struxture makes FMO far from
ideal as a first example to study. We thus prefer to start by working
with a much simpler, yet fully relevant subsystem: the trimeric
unit composed by the sites $1,2$ and $3$ of the FMO complex in the
notation of~\cite{Renger,Cho}). The first chromophore is the site
in which the energy transfer begins, while the third chromophore is
the site from which the excitation leaves the complex. The Hamiltonian
of the trimeric subunit is \cite{Renger} 
\begin{equation}
H_{Renger}=\left(\begin{array}{ccc}
215 & -104.1 & 5.1\\
-104.1 & 220 & 32.6\\
5.1 & 32.6 & 0
\end{array}\right).\label{eq: HamTrimer Renger}
\end{equation}
The eigenenergies of the system ar given by\textbf{ $E_{+}=322.85\,cm^{-1},$$E_{-}=119.13\,cm^{-1},$
}$\,E_{3}=-6.98\,cm^{-1}$ which yields the eigenperiods
$T_{ij}=(2\pi\hbar/\Delta E_{ij}):$ $T_{+-}=0.163\,ps,\,T_{-3}=0.100\,ps,\,T_{+3}=0.264\,ps$.
Due its structure, the trimer is a chain composed by a pair of chromophores
($1,2$), degenerate in energy and forming a strongly coupled dimer,
and a third chromophore moderately coupled with the second one only.
Since in the following we suppose that the exciton starts from site
$1$, we expect a prominent role of the dimer in the dynamics, at least
in the first tens of femtoseconds.\\

\begin{figure*}[tbh]
\subfigure[]{\includegraphics[width=0.4\textwidth]{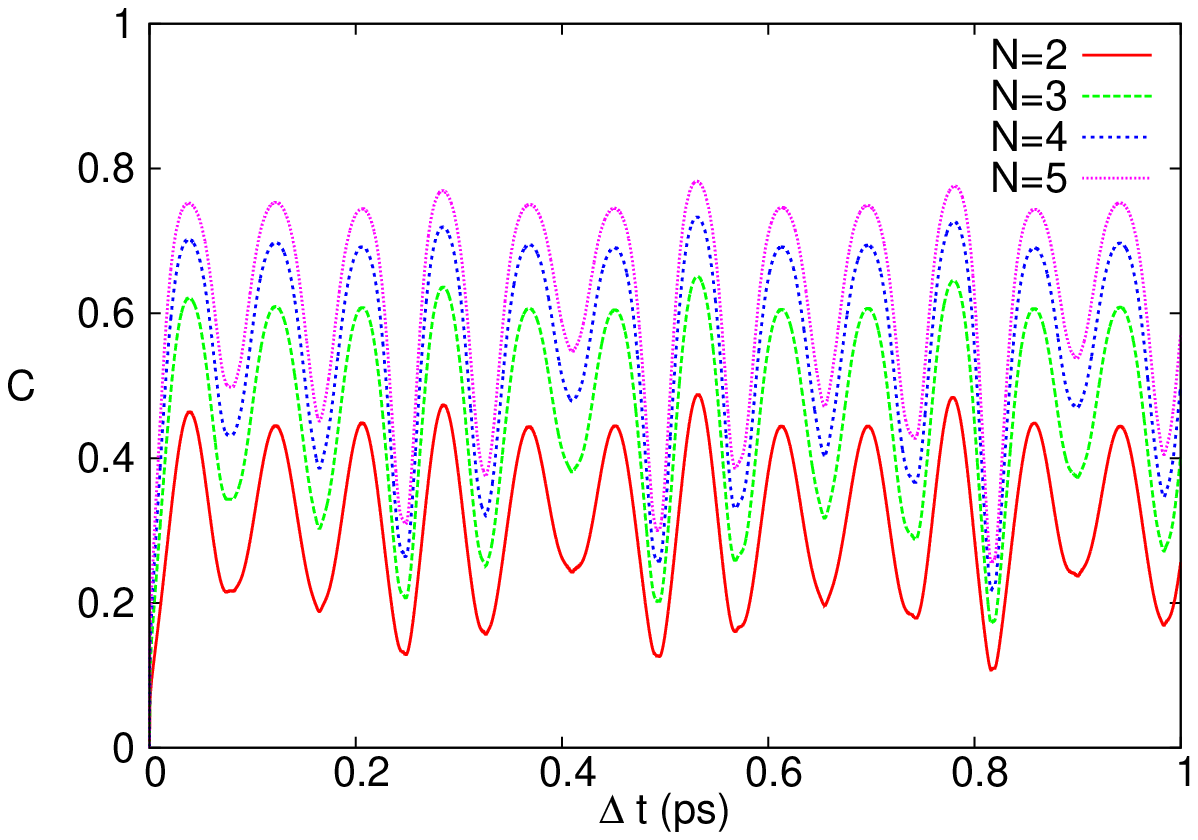}}
\subfigure[]{\includegraphics[width=0.4\textwidth]{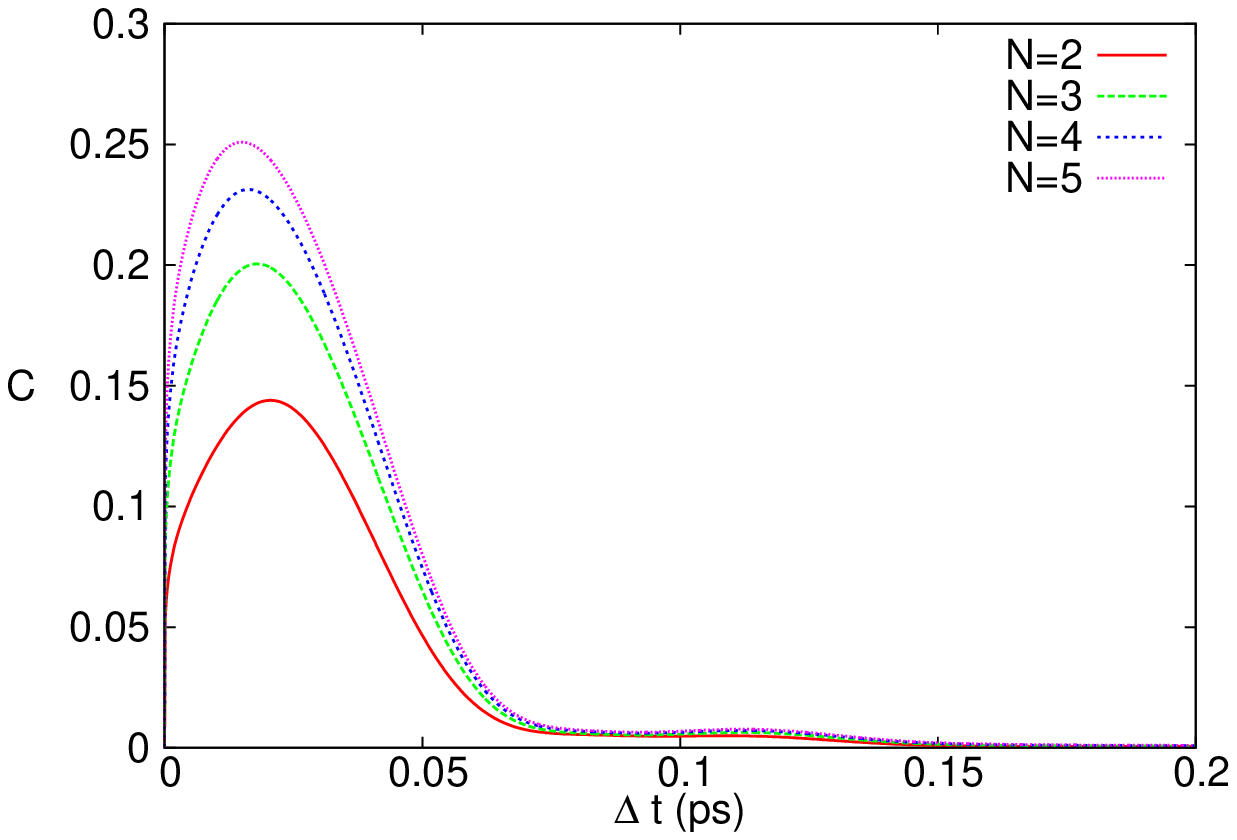}}
\subfigure[]{\includegraphics[width=0.4\textwidth]{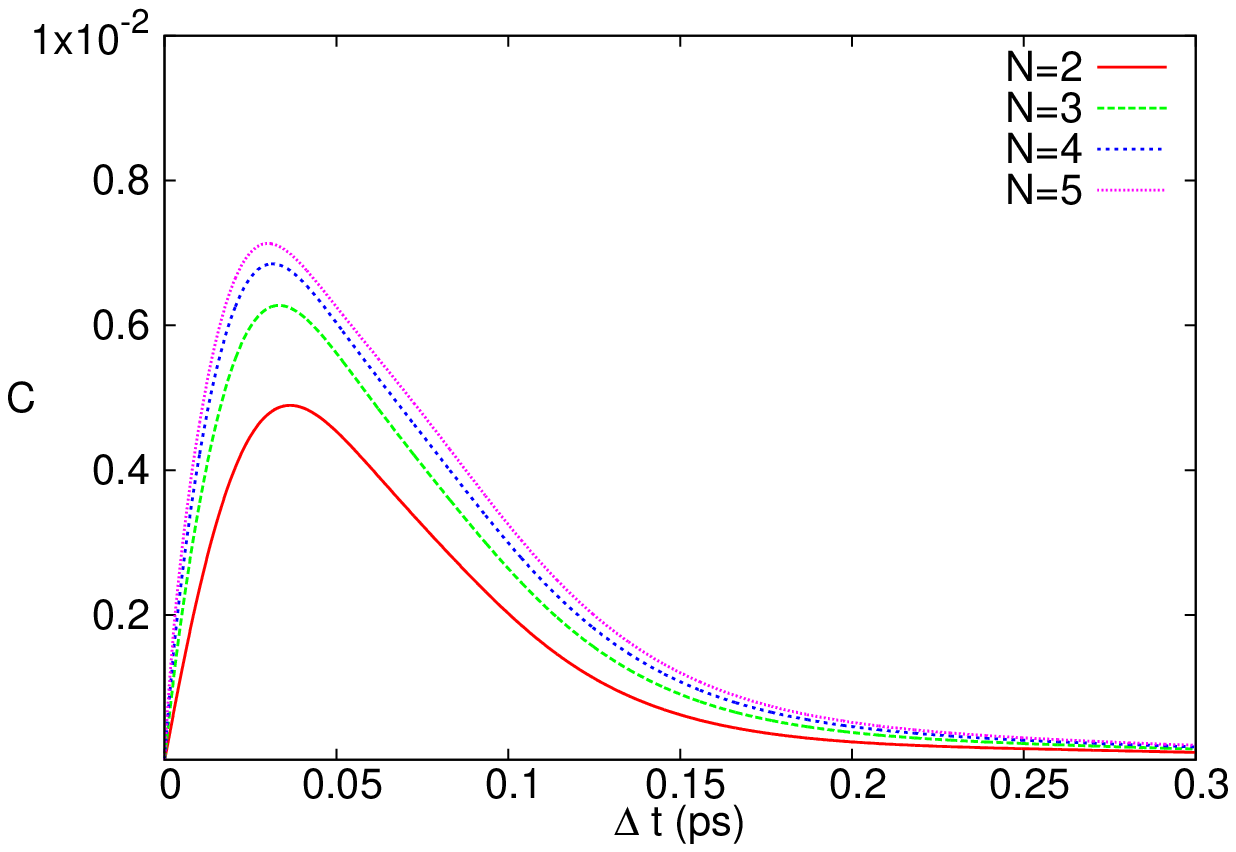}}
\subfigure[]{\includegraphics[width=0.4\textwidth]{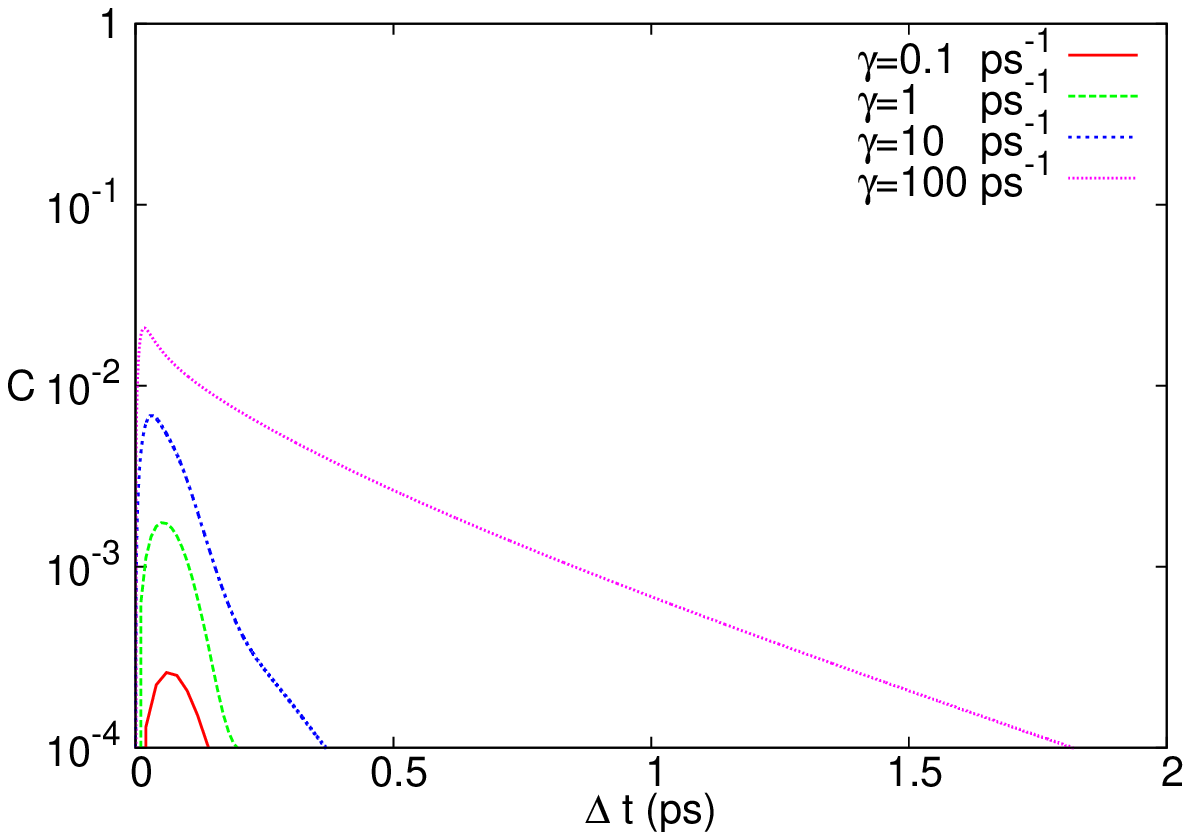}}\protect\caption{Coherence function $C(P,N,\Delta t)$ as a function of $\Delta t$
for the trimer with Hamiltonian \ref{eq: HamTrimer Renger} evaluated:
\textbf{(a)} in the site basis for $\gamma=0$, \textbf{(b)} in the site basis for $\gamma=10 ps^{-1}$ \textbf{(c)} in the exciton basis
for $\gamma=10 ps^{-1}$ \textbf{(d)} in the exciton basis for different values of dephasing $\gamma$. }
\label{Fig: Trimer Renger C function} 
\end{figure*}

In order to show how the DH analysis can be implemented,
in the following we are going to consider histories in the site and the energy bases,
with $N$ projections at times $n\Delta t,\:n=1,..,N$. We first
use the coherence function $C(P,N,\Delta t)$ introduced above (\ref{eq: Cfunction})
to evaluate the global coherence of the exciton transport process.
In order to test the behavior of $C$ for different values of dephasing,
in Fig \ref{Fig: Trimer Renger C function} we first plot $C$ as
a function of the time interval $\Delta t$ between projections for two values
of the dephasing rate: $i)\,\gamma=0$, corresponding to the full
quantum regime (Fig.\ref{Fig: Trimer Renger C function}) for the
site basis (a); $ii)\:\gamma=10$ corresponding to an intermediate
value of dephasing (Fig. \ref{Fig: Trimer Renger C function}) for
the site basis (b) and the energy basis (c).

Before entering the discussion of the various regimes, we
note that as a function of the number of projections $N$ all curves
display the expected behavior: The increase (decrease) of the number
of projections corresponds to a temporal fine-graining (coarse-graining)
of the evolution; therefore, an increase (decrease) of $N$ should
imply an increase (decrease) of the amount of coherence between histories.
As shown in Fig. (\ref{Fig: Trimer Renger C function}) the function
$C$ correctly reproduces the fine (coarse) graining feature: the
qualitative behavior of $C$ as a function of $\Delta t$ is not
affected by the choice of $N$, while an increase of $N$ corresponds,
at fixed $\Delta t$, to an increase of $C$. 
We will therefore use in the following the value $N=4$ that allows
for a neat description of the phenomena and for a reasonable computational
time.

As for the behavior at fixed $N$, we have that in the full quantum
regime ($\gamma=0$), the system obviously displays coherence in the
site basis only since 
\[
\Tr[\ket{E_{i}}\bra{E_{i}}e^{-iH\Delta t}\rho e^{iH\Delta t}\ket{E_{j}}\bra{E_{j}}]=\bra{E_{i}}\rho\ket{E_{i}}\delta_{i,j}
\]
$ $and the decoherence matrix $\DD$ in the energy basis is diagonal
and independent on $\Delta t$ and $N$. This simply means that in
the full quantum regime histories in the exciton basis are fully decohered,\textbf{
}since the system is not able to create coherence among excitons.
Still in the full quantum regime, in the site basis, the coherence oscillates
as the exciton, starting at site $1$, goes back an forth along the
trimer, and the evolution builds up coherence in this basis, see Fig.\ref{Fig: Trimer Renger C function}
(a). In this regime, the trimer can be approximately seen as a dimer
composed by the first two chromophores, and the exciton performs
Rabi oscillations with a period given by $T_{+-}=160\,fs$; $C$
oscillates with half the period: for $\Delta t=80\,fs$ the exciton
is migrated mostly on site $2$ and $C$ has a minumum -- which is different
from zero since the exciton is partly delocalized on site $3$, and
the system therefore exhibits a non vanishing coherence.

For intermediate values of $\gamma\approx10\,ps^{-1}$ , Fig.\ref{Fig: Trimer Renger C function}
(b), the coherence in site basis as measured by $C$ correctly drops
down at $\Delta t\ge\gamma^{-1}$\cite{ZurekHist}. The dephasing
has a strong and obvious effect on the coherence between pathways:
coherence in this basis is a monotonically decreasing function of $\gamma$.
This is well highlighted by the global coherence function $C$, whose
maximal values are reduced by a factor of $\sim 3$ with respect to
those corresponding to full quantum regime. After a time $\tau_{decoh}=\gamma^{-1}$
the histories are fully decohered. Indeed, due to the specific model
of decoherence (\ref{eq: HakenStrobl}), which amounts to projective
measurements on $\ket{i}\bra{i}$ at each site with a rate $\gamma$,
the system kills the coherence in the site basis, which in turn corresponds
to the stable pointer basis for this model \cite{ZurekHist}, i.e., the
basis in which the density matrix is forced to be diagonal
by the specific decoherence model. On the other hand,
and for the same reason, the dynamics starts to build up coherence
in the exciton basis $\ket{E_{i}}\bra{E_{i}}$, see Fig. \ref{Fig: Trimer Renger C function}(c).
However, this coherence is later destroyed - on a time scale of approximately $0.2\:ps$ - since the
stationary state of the model is the identity. This effect is even more
evident if one compares the behavior of $C$ in the exciton basis for
different values of $\gamma$, as shown in Fig. \ref{Fig: Trimer Renger C function}(d):
$C$ grows with $\gamma$ and it lasts over longer time scales. This
feature is coherent with the expectations: the equilibrium state for
high $\gamma$ is the identity. Due to the projections implemented
by the environment in the site basis, the system is forced to create coherence
in the exciton basis. When $\gamma$ is very high a quantum Zeno effect
takes in, the dynamics is blocked, and the time required to reach
the equilibrium, and to destroy coherences in all bases, consequently grows.

This first analysis therefore shows that $C$ is indeed a good candidate
for assessing the global coherence properties of quantum evolutions.
For a fixed number of projections $N$, $C(\Delta t)$ can be interpreted
as a \emph{measure of the global coherence exhibited by the dynamics
over the time scale} $\Delta t$.

%characterized by the Hamiltonian
%We first consider a trimer, i.e., 3-site network. %characterized by the Hamiltonian
%This is virtually the simplest energy transfer network that can be studied. 
%This is virtually the simplest energy transfer network that can be studied. 
%The $\Delta < 0 $ accounts for an effect of energy funnel towards the target site $\bar{x}=3$. 
%The initial state of the exciton is $|1 \rangle$. \\
%We simulate different dephasing conditions: i) no dephasing is present, $\gamma_i=0$ ii) the dephasing is the uniform globally optimized dephasing, $\gamma_i=\gamma_{glob}^{opt}$ iii)  the dephasing is the locally optimized dephasing, $\gamma_i=\gamma_{loc, i}^{opt}$. \\
%The globally and locally optimized dephasing are those that minimize the
%the trapping tim $\tau$, defined as follows,
%\begin{displaymath}
%\tau = 2k_{trap} \int_0^\infty dt \ t \ \mbox{Tr}  [|3\rangle \langle 3| \varrho(t)]
%\end{displaymath}
%If we fix $k_{trap}=5$ and we consider a globally uniform dephasing
%$\gamma_i = \gamma$, $\forall i$, then $\tau$ becomes a function of $\gamma$. Its behavior is plotted in Fig.~\ref{Figure: trapping3sites}a. The minimum $\tau$ is achieved for a value $\gamma_{glob}^{opt}=10.76$, leading to
%$\tau=0.554$. If we allow for locally nonuniform values of dephasing, we can further optimize $\tau$. In particular, we can take
%the dephasing values $\gamma_{loc,i}^{opt}=\{10,20,0\}$, leading to $\tau=0.520$. \\

\begin{figure}[t]
\subfigure[]{\includegraphics[width=0.4\textwidth]{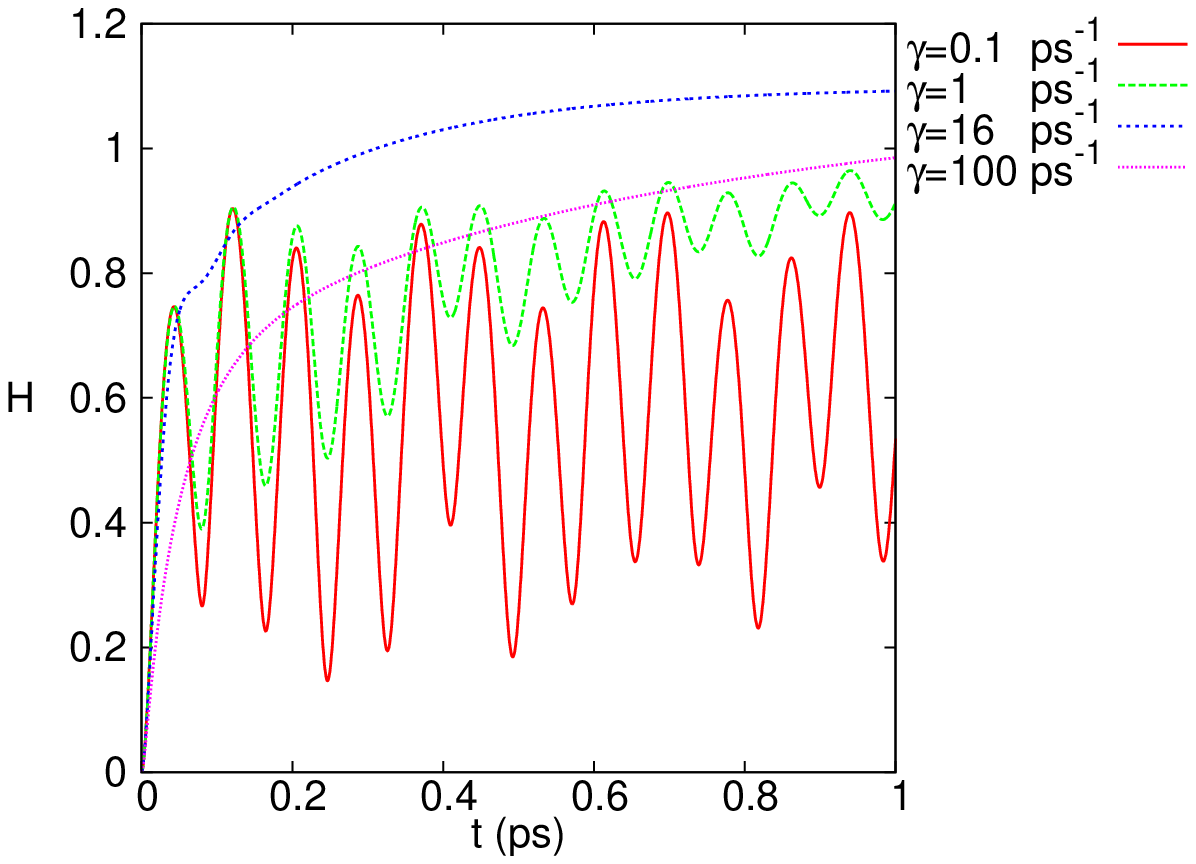}}
\subfigure[]{\includegraphics[width=0.4\textwidth]{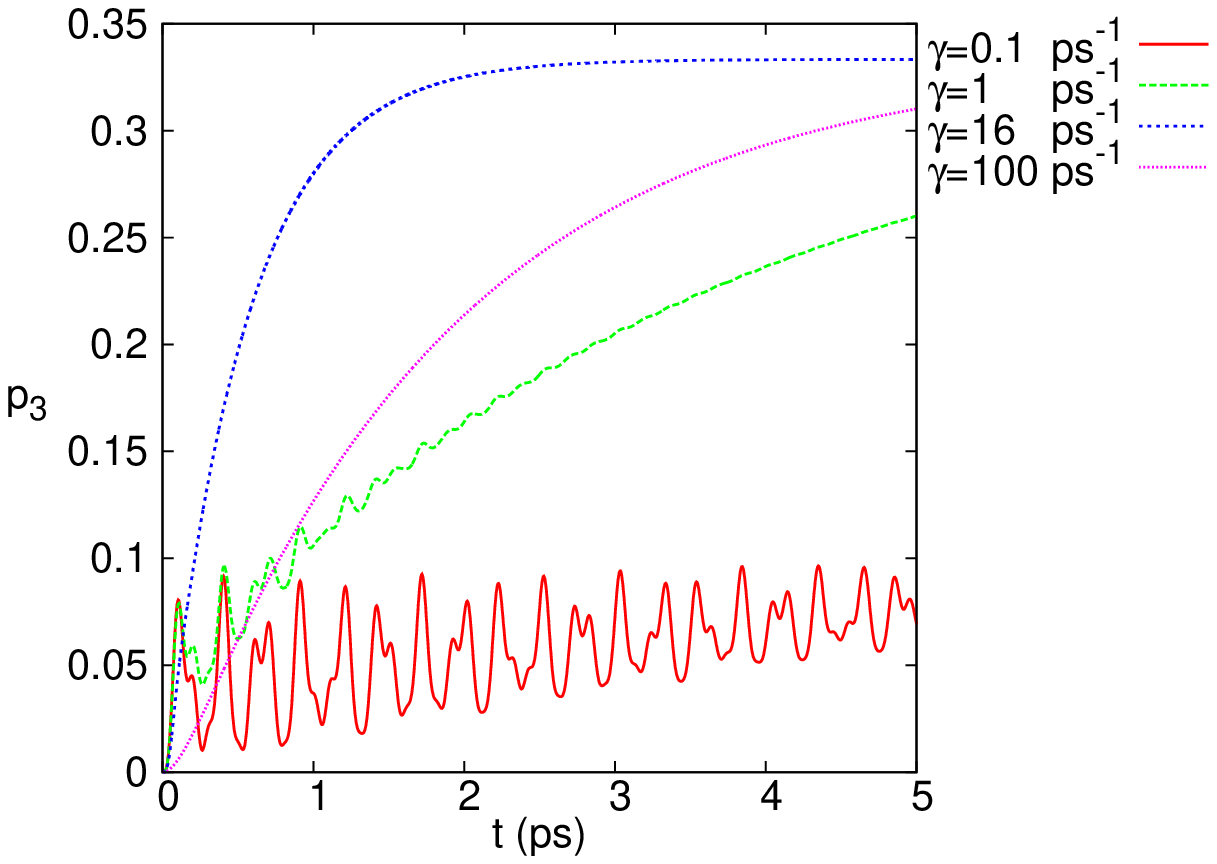}}
\protect\caption{For the trimer (\ref{eq: HamTrimer Renger}): \textbf{(a)} Delocalization
$\mathcal{{H}}(t)$ and \textbf{(b)} $p_{3}(t)$ population of site $3$
as a function of time for different values of $\gamma=0.1,1,16,100$ $ps^{-1}$}

\label{Fig.: pop site 3 and delocalization} 
\end{figure}

We now analyze in detail the specific features of quantum transport
for the trimer. The dynamics starts at site $1$ and evolves by
delocalizing the exciton on the other chromophores. In order to study
this process, we first use a measure of delocalization introduced in\cite{GiordaLHCII}
for the study of LHCII complex dynamics: 
\begin{equation}
\mathcal{{H}}(t)=-\sum_{i}p_{i}(t)\ln p_{i}(t)
\end{equation}
that is simply the Shannon entropy of $p_{i}(t)$, the populations of the three
chromophores. This measure allows one to follow how much the exciton
gets delocalized over the trimer with time and in different dephasing
situations: $0\le\mathcal{{H}}(t)\le\ln(3)$, i.e., $\mathcal{{H}}$
is zero when the exciton is localized on a chromophore and it takes
its maximal value $\ln(3)$ when the population of the three sites are equal.
In Fig. \ref{Fig.: pop site 3 and delocalization} we plot both $\mathcal{{H}}(t)$
and the population $p_{3}(t)$ of site $3$ for different values of
$\gamma$. Due to the presence of interference, in the mainly quantum
regime ($\gamma=0.1,1\:ps^{-1}$), the exiton first delocalizes mainly
over the dimer and partly on the third site: The first maximum corresponds
to $t=40\:fs=1/4 \ T_{+-} $ when the system builds up a (close to
uniform) coherent superposition between sites $1$ and $2$, while
a non negligible part of the exciton is found in site $3$; indeed
$\mathcal{{H}}(t=40\,fs)\approx0.75>0.69$, the last value corresponding
to $\ln(2)$ i.e., to a uniform superposition over the sites $1$
and $2$ only. As the dynamics of the sytems extends to later times
we see that $\mathcal{{H}}(t)$ and $p_{3}(t)$
have an oscillatory behavior, whose main period is $1/2 \ T_{+-}$, 
and which approximately corresponds to Rabi oscillations between site
$1$ and $2$, although the initial state fully localized in site
$1$ cannot be rebuilt due to the presence of site $3$. As for the
transport, we see that in this regime the system cannot take advantage
of the initial fast and high delocalization: the exciton bounces back
and forth over the trimer. In the intermediate regime $\gamma=16\:ps^{-1}$,
due, as we will later see, to the selective suppression of interference
processes, the initial speed up in delocalization is sustained by
the dynamical evolution, and the transfer rate to site $3$ is correspondingly
increased. For very high values of decoherence ($\gamma=100$) the
role of initial interference is suppressed and the initial speed-up
disappears: the environment measures the system in site basis at high
rates and the delocalization process is highly reduced.

The optimal delocalization occurs in correspondence of $\gamma\approx16\:ps^{-1}$
and it can be interpolated with a double exponential function

\[
\mathcal{{H}}_{\gamma=16}(t)=c_{0}+c_{1}e^{-t/\tau_{1}}+c_{2}e^{-t/\tau_{2}}
\]
with $c_{0}=1.098=\ln3,\,c_{1}=-0.84,\,c_{2}=-0.373$. The first time
scale $\tau_{1}=23\:fs$ describes the initial fast quantum delocalization
process described above, while the second time scale $\tau_{2}=238\:fs$
the slower subsequent delocalization and the reaching of the equilibrium
situation, $\mathcal{{H}}(t=\infty)=\ln(3)$.

\begin{figure}[t]
\includegraphics[width=0.4\textwidth]{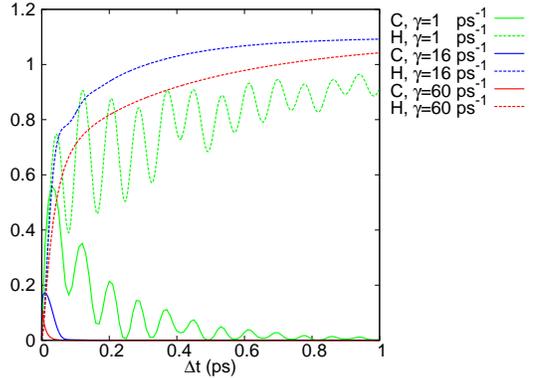}\protect\caption{Delocalization $\mathcal{{H}}(\tau=N\Delta t)$ and coherence function
$\mathcal{{C}}(\Delta t)$ for $\gamma=1,16,60$ $ps^{-1}$}
\label{Fig.: H and C various gamma} 
\end{figure}

We now pass to systematically analyze the behavior of the coherence
of the evolution with respect to the strength of the interaction with
the environment and its relevance for the energy transport process.
As a first step we plot both $\mathcal{{H}}(\tau=N\Delta t)$ and
$\mathcal{{C}}(\Delta t)$ for different values of $\gamma$, Fig.
(\ref{Fig.: H and C various gamma}). The plots show that the coherence
function exhibits the required behavior: For small $\gamma=0.1$
, $\mathcal{{C}}(\Delta t)$ oscillates with period $1/2 \ T_{+-}$,
following the Rabi oscillations of the dimer. The minima occur at	
$n/4 \ T_{+-}$ , showing that the exciton is ``partially'' localized
on site $1$ or $2$, and partially delocalized on site $3$. As $\gamma$
grows, the system becomes unable to create coherence on large time
scales; the decay of $\mathcal{{C}}(\Delta t)$ is mirrored by a the
reduction of the amplitude in the oscillations of $\mathcal{{H}}(\tau=N\Delta t)$.

\begin{figure}[t]
\includegraphics[width=0.4\textwidth]{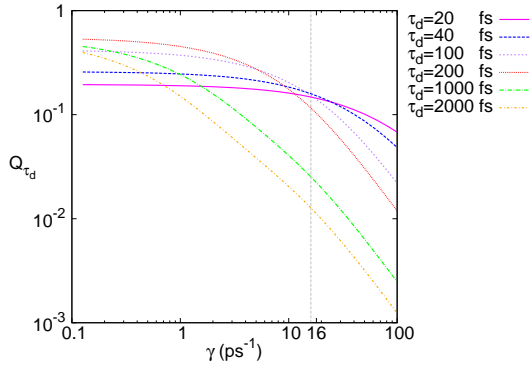}\protect\caption{Average coherence of the evolution $Q_{\tau_{d}}(\gamma)$ for the
trimer (\ref{eq: HamTrimer Renger}) for different values of $\tau_{d}$}

\label{Fig.: Average C vs gamma} 
\end{figure}

We now focus on the relevant time scales $\tau_{d}$ for the initial
fast delocalization process highlighted by our previous analysis,
which are of the order of tens to hundreds of femtoseconds. We
therefore introduce the following \emph{average measure of global
coherence of the evolution} 
\[
Q_{\tau_{d}}(\gamma)=\frac{1}{\tau_{d}}\int_{0}^{\tau_{d}}C(\Delta t)\,d\Delta t.
\]
$Q_{\tau_{d}}(\gamma)$ is the average of the coherence exhibited
by the dynamics of the system at the time scales $\Delta t\in(0,\tau_{d})$.
In Fig. \ref{Fig.: Average C vs gamma} we show $Q_{\tau_{d}}(\gamma)$
for the trimer (\ref{eq: HamTrimer Renger}) in the site basis for different
values of $\tau_{d}$.\textbf{ }We first focus on the behavior of
$Q_{\tau_{d}}(\gamma)$ for values of dephasing in the range $\gamma\in(0,1)$ $ps^{-1}$.
In this range, for small time scales $\tau_{d}=20$ to $200\,fs$
the average global coherence $Q_{\tau_{d}}(\gamma)$ is approximately
constant and equals the value attained in the full quantum regime
i.e., $Q_{\tau_{d}}(\gamma)\approx Q_{\tau_{d}}(\gamma=0.1)$ . For
larger time scales ($\tau_{d}\approx1\:ps$) $Q_{\tau_{d}}(\gamma)$ rapidly decreases
with $\gamma$. This analysis shows that the behavior of $Q_{\tau_{d}}(\gamma)$
matches the expectations: the higher $\gamma$ the smaller the time
scales over which decoherence takes place, the lower the global coherence
of the dynamics. Along with $C(\Delta t)$ the functional $Q_{\tau_{d}}(\gamma)$
is therefore in general a good candidate for the evaluation of the
global coherence of open quantum systems evolution. As for the transport
dynamics, we focus on the timescale identified with the analysis of
$\mathcal{{H}}(t)$ for optimal dephasing; for $\tau_{d}=\tau_{1}=20\,ps$
and $\tau_{d}=40\,ps$ we see that the system indeed retains most
of the average coherence of the purely quantum regime up to the optimal
values of decoherence ($\gamma=16\:ps^{-1}$ in the figure), losing it afterwards; this is a clear indication that this phenomenon
is at the basis of the the fast initial delocalization process. Over
longer time scales, the relevance of coherence is highly suppressed\textbf{.}

We now deepen our analysis about the relevance of the coherence of
the evolution for the energy tranfer efficiency. To this aim we focus
on the basic feature that distingushes the classical and the quantum
regime: interference. In particular we focus on the sub-block $\DD_{3}$
of the decoherence matrix $\DD$ pertaining to the third chromophore,
which describes the set of histories in site basis ending at site
$3$. Due to interference the probability of occupation of the site
$3$ at time $\tau=N\Delta t$ can be written in terms of the the
histories ending at site $3$ $p_{3}(\tau)=w_{3}(\tau)+\II_{3}(\mathbf{\tau})$,
see (\ref{eq: interference and probability at ending site j}) .

\begin{figure}[t]
\subfigure[]{\includegraphics[width=0.4\textwidth]{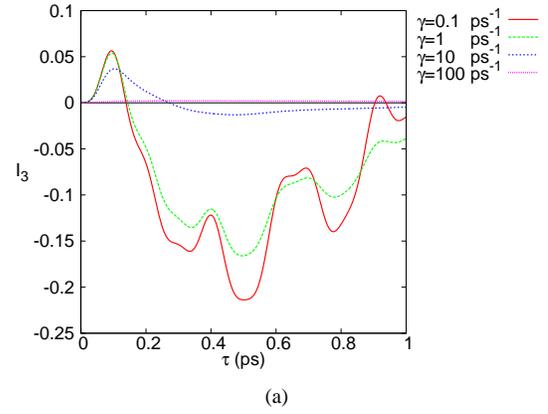}}
\subfigure[]{\includegraphics[width=0.4\textwidth]{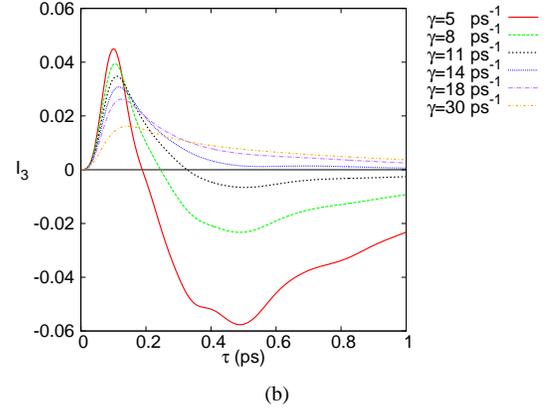}}
\protect\caption{\textbf{(a)} interference $\mathcal{{I}}_{3}(\tau)$ of histories ending
in site $3$ for the trimer (\ref{eq: HamTrimer Renger}) as a function
of $\tau=N\Delta t$ for different values of $\gamma=0.1,1,10,100$;
\textbf{(b)} $\mathcal{{I}}_{3}(\tau)$ for different intermediate values
of $\gamma$}

\label{Fig.: Interference with gamma Trimer} 
\end{figure}

In Fig. \ref{Fig.: Interference with gamma Trimer} we show $\II_{3}(\tau)$
for different values of dephasing. One has different regimes: for
$\gamma\gg1$, the set of histories in site basis is fully decohered;
$\II_{3}(\tau)\approx0$, the histories do not interfere with each
other and $p_{3}(\tau)\approx w_{3}(\tau)$ i.e, the probability is
simply the sum of diagonal elements of $\DD_{3}$. In the mainly quantum
regime $\gamma\le1\,ps^{-1}$, $p_{3}(\tau)\neq w_{3}(\tau)$: after
the initial positive peak the histories interfere with each other,
globally the interference is mostly negative and therefore $p_{3}(\tau)\le w_{3}(\tau)$.
For intermediate values of decoherence $\gamma^{-1}\approx10\:ps$
the interference has a positive peak and then reduces to zero. While
the first initial fingersnap of positive interference that takes place
in the first $\approx80\:fs$ is common for all curves corresponding
to small and intermediate values of $\gamma$, the \emph{main effect
of the bath is displayed after this initial period of time: the decoherence
gradually suppresses interference, both the positive and the negative
one}; \emph{however, for intermediate values of $\gamma$ the effect
is stronger as for the negative part of the interference patterns}.
The environment thus implements what can be called a \emph{quantum
recoil avoiding effect:} it prevents the part of the exciton that --
thanks to constructive interference -- has delocalized on site $3$
to flow back to the the other sites.

\begin{figure}[t]
\includegraphics[width=0.4\textwidth]{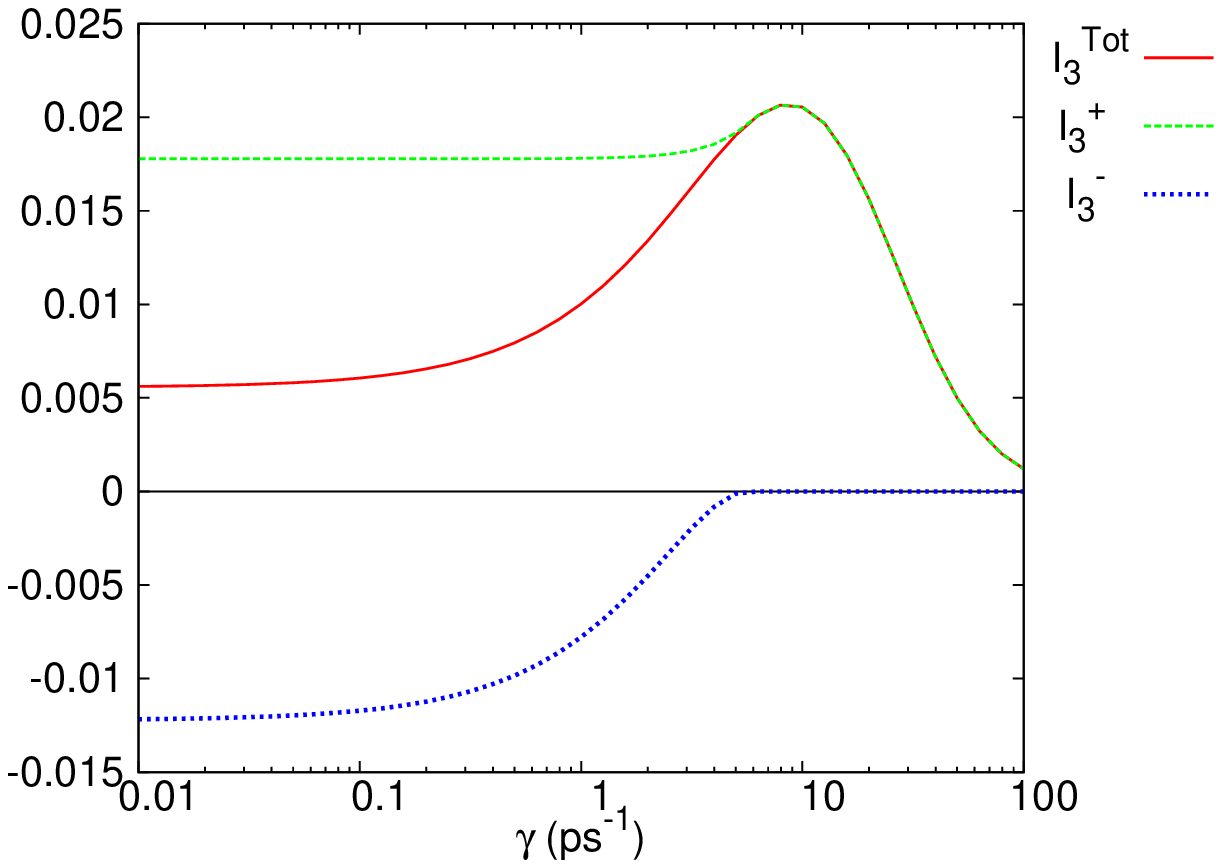}
\includegraphics[width=0.4\textwidth]{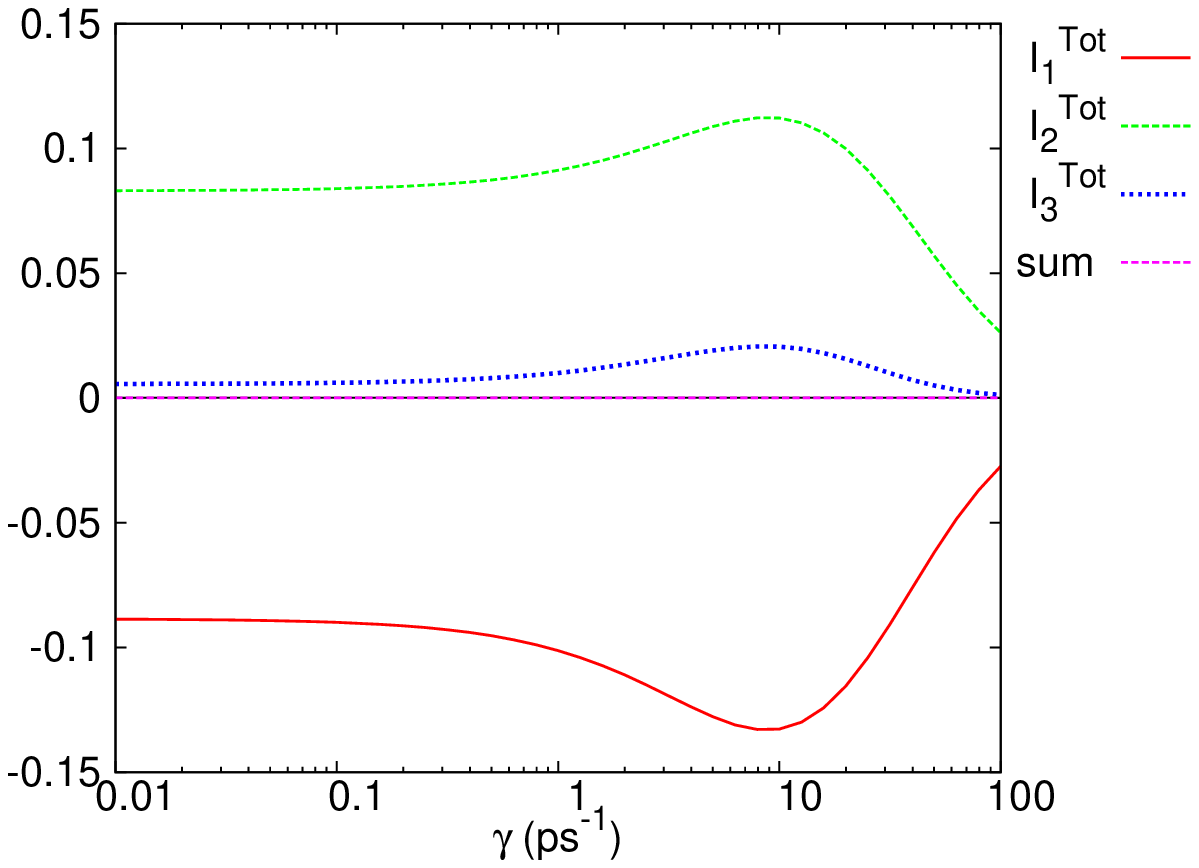}
\protect\caption{For the trimer (\ref{eq: HamTrimer Renger}) \textbf{(a)} Average
positive (green), negative (blue), total (red) interference $\average{\mathcal{{I}}_{3}}$
of histories ending in site $3$ as a function of $\gamma$; \textbf{(b)}
total average interference $\average{\mathcal{{I}}_{i}^{Tot}}$ for
site $i=1$ (red), $i=2$ (blue), $i=3$ (green) $\gamma$.}

\label{Fig.: average interference site 3} 
\end{figure}

In order to evaluate a possible advantage provided by the initial speed-up 
in the delocalization process and by the interference phenomena
showed above one has to take into account another relevant time scale
of the transport process: the trapping time. Indeed, if the system
is to take advantage of the fast delocalization due to the coherent
behavior, the exit of the exciton should take place on time scales
of the order of the delocalization process. The theoretical and experimental
evidences show that this is the case: the trapping time $\tau_{trap}=k_{trap}^{-1}$
for the FMO complex is estimated in the literature to be of the order
of $0.2\:ps$ i.e., the exit of the exiton starts soon after the fast
delocalization due to quantum coherence has taken place. The role
of the interference between paths, in particular those leading to
site $3$, can therefore be appreciated by numerically evaluating
\begin{equation}
\average{\II_{i}^{\beta}}=\frac{{1}}{\tau_{trap}}\int_{0}^{\tau_{trap}}\II_{i}^{\beta}(\tau)d\tau
\end{equation}
i.e., the average over the trapping time scale of $\tau_{trap}=200\:fs$
of the total ($\beta=Tot$), negative ($\beta=-$) and positive ($\beta=+$)$ $
average interference between the histories ending in site $i$, with
$\average{\II_{i}^{Tot}}=\average{\II_{i}^{+}}+\average{\II_{i}^{-}}$.
In particular, in Fig. \ref{Fig.: average interference site 3}(a)
the different kinds of interference are plotted for histories terminating
at site $3$: on average, the negative interference highly reduces
the total interference for small values of decoherence strength; when
$\gamma\approx10\:ps^{-1}$, $\average{\II_{3}^{-}}$ vanishes, the
average total interference equals the positive one $\average{\II_{3}^{Tot}}=\average{\II_{3}^{+}}$,	
and it is maximal for values of $\gamma$ comparable to those that
maximize $\mathcal{{H}}(t)$ ($\approx16\:ps^{-1}$). In Fig. \ref{Fig.: average interference site 3}(b), 
we compare the behavior of $\average{\II_{i}^{Tot}}$ for
all sites. The results again suggest that decoherence acts on the
interference provided by the quantum engine in order to favor the
flow of the exciton towards the exit chromophore: the average positive
interference between histories ending at sites $2$ and $3$ grows
in modulus with $\gamma$ and attains a maximum for intermediate values
of decoherence; while the average negative interference between histories
ending at site 1 decreases and attains a minimum for intermediate values
of $\gamma$. The combined effect of decoherence and interference
thus helps depopulating site 1 and populating site 2 and 3.

\begin{figure}[t]
\includegraphics[width=0.4\textwidth]{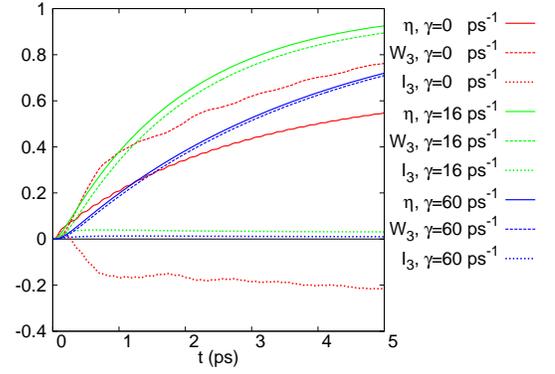}
\protect\caption{For the trimer (\ref{eq: HamTrimer Renger}): transport efficiency
$\eta(t)$, integrated weight $W_{3}(t)=2k_{trap}\intop_{0}^{t}w_{3}(\tau)d\tau$ and
integrated interference $I_{3}(t)=2k_{trap}\intop_{0}^{t}\mathcal{{I}}_{3}(\tau)d\tau$ for pathways
ending at site $3$ for different values of $\gamma=0.1,16,60$ $ps^{-1}$ and
$k_{trap}=5\,ps^{-1}$}
\label{Fig.: efficiency vs interference trimer} 
\end{figure}

We can now tackle one of the most relevant aspects of our discussion:
the net effect of the above described phenomena on the overall efficiency
of the transport. The latter can be fully appreciated by evaluating
the efficiency of the process (\ref{eq: efficiency}) and by recognizing
that, in the decoherent histories language, it can be expressed as:

\[
\eta(t)=2k_{trap}\intop_{0}^{t}p_{3}(\tau)d\tau=W_{3}(t)+I_{3}(t)
\]
where $\tau=N\Delta t$ and $W_{3}(t)=2k_{trap}\intop_{0}^{t}w_{3}(\tau)d\tau,\ I_{3}(t)=2k_{trap}\intop_{0}^{t}\mathcal{{I}}_{3}(\tau)d\tau$.
This split allows one to appriciate the role of interference for the
efficiency. In Fig. (\ref{Fig.: efficiency vs interference trimer})
$\eta$ is plotted for different values of dephasing. In agreement
with what discussed above, we have three regimes: for very small values
of $\gamma$ the overall efficiency is poor; this is due to the presence
of high negative interference that in average prevents the exciton
to migrate to the exit site. For large values of $\gamma$ the interference
processes are completely washed out and the system cannot take advantage
of the fast quantum delocalization. For intermediate (optimal) values
of $\gamma$ only the negative interference has been washed out: $\mathcal{{I}}_{3}(\tau)$
is positive, it acts on short time scales, and it provides on average
an enhancement of the global efficiency.

These results, within the limits of the simple model of decoherence
taken into account, undoubtedly show for the first time that the so
called ENAQT phenomenon can well and properly be understood both qualitatively
and quantitatively within the decoherence histories approach, i.e.,
in terms of very the basic concepts of coherence and interference
between histories. The often recalled ``convergence'' of time scales
or ``Goldilocks'' effect (\cite{Goldilocks}) in biological quantum
transport systems seems therefore to be well rooted in the processes
discussed above: if decoherence is too small the system shows both
positive and negative interference (see Fig. \ref{Fig.: pop site 3 and delocalization}),
the delocalization has an ocillatory behavior, and the exciton bounces
back and forth along the network thus preventing its efficient extraction.
If instead decoherence is very high one has that the complete washing
out of intereference and coherence implies the delocalization process
to be very slow, no matter how fast the trapping mechanism try to
suck the exciton out of the system. In order to take advantage of
the effects of quantum coherent dynamics: $i)$ the bath must act
on the typical time scales of quantum evolution in order to implement
the quantum recoil avoiding process; $ii)$ the extraction of the
exciton from the complex, characterized by $k_{trap}$, must then
start soon after the initial fast delocalization has taken place.
Should the extraction take place on longer time scales, the benefits
of the fast initial delocalization would be spoiled: Waiting
long enough, the system would eventually reach together with equilibrium
a decent delocalization even for moderately high values of $\gamma$,
but in this case the transfer would be obviously much slower.

\section{FMO\label{sec:FMO}}

\begin{figure}[t]
\subfigure[]{\includegraphics[width=0.4\textwidth]{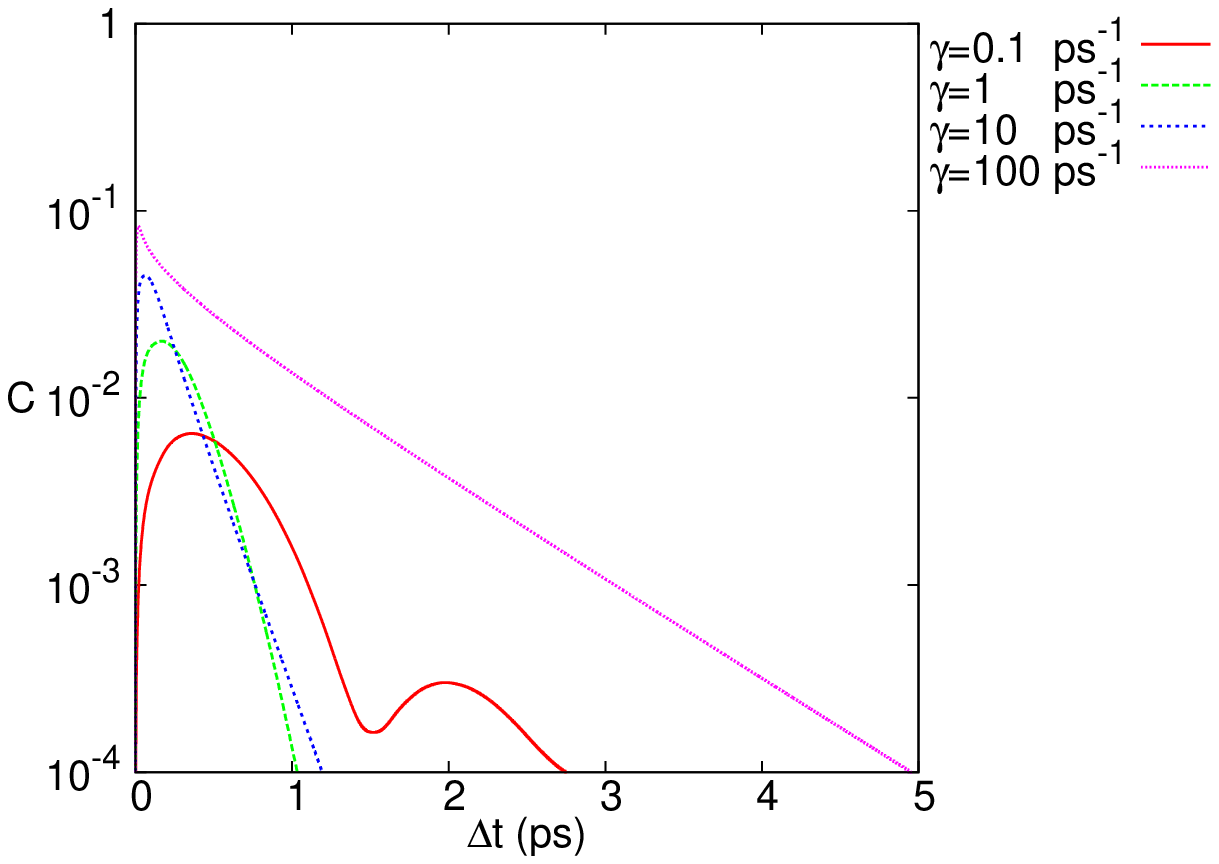}}
\subfigure[]{\includegraphics[width=0.4\textwidth]{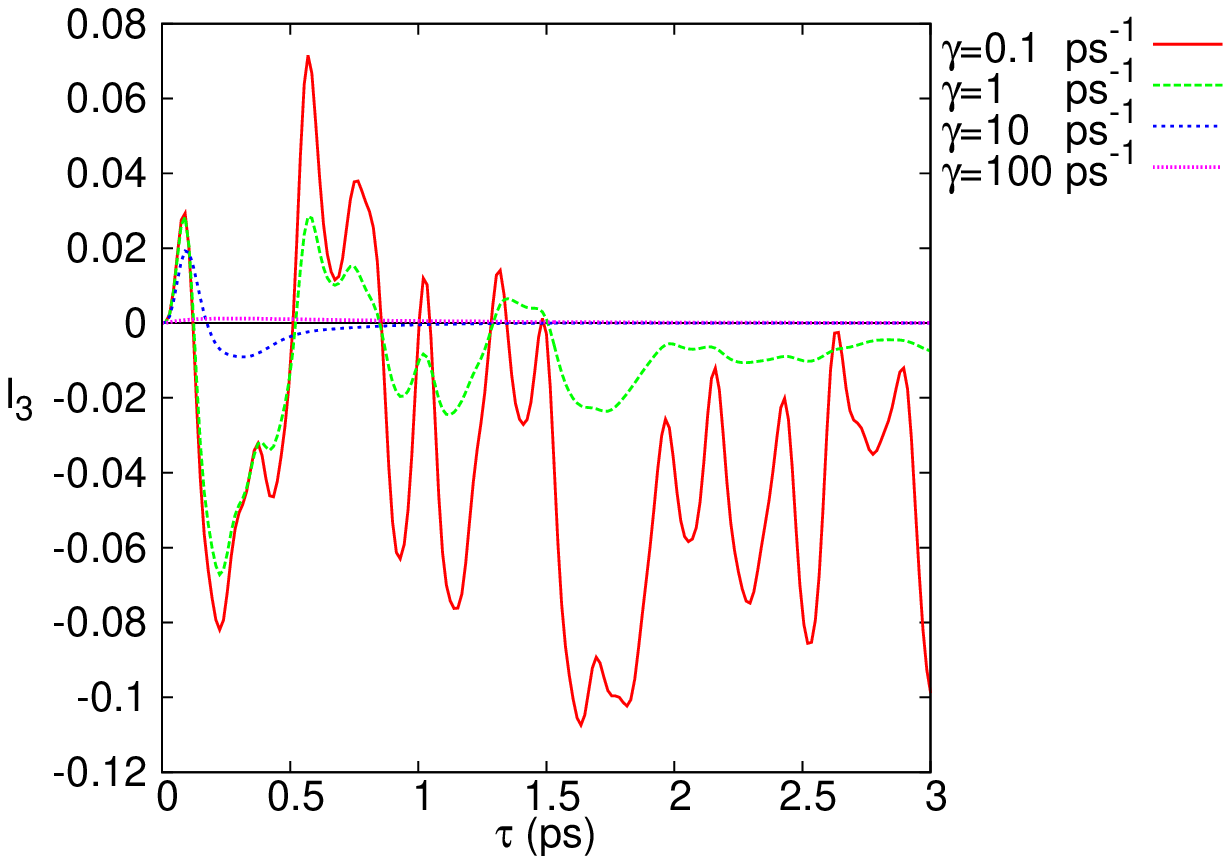}}
\subfigure[]{\includegraphics[width=0.4\textwidth]{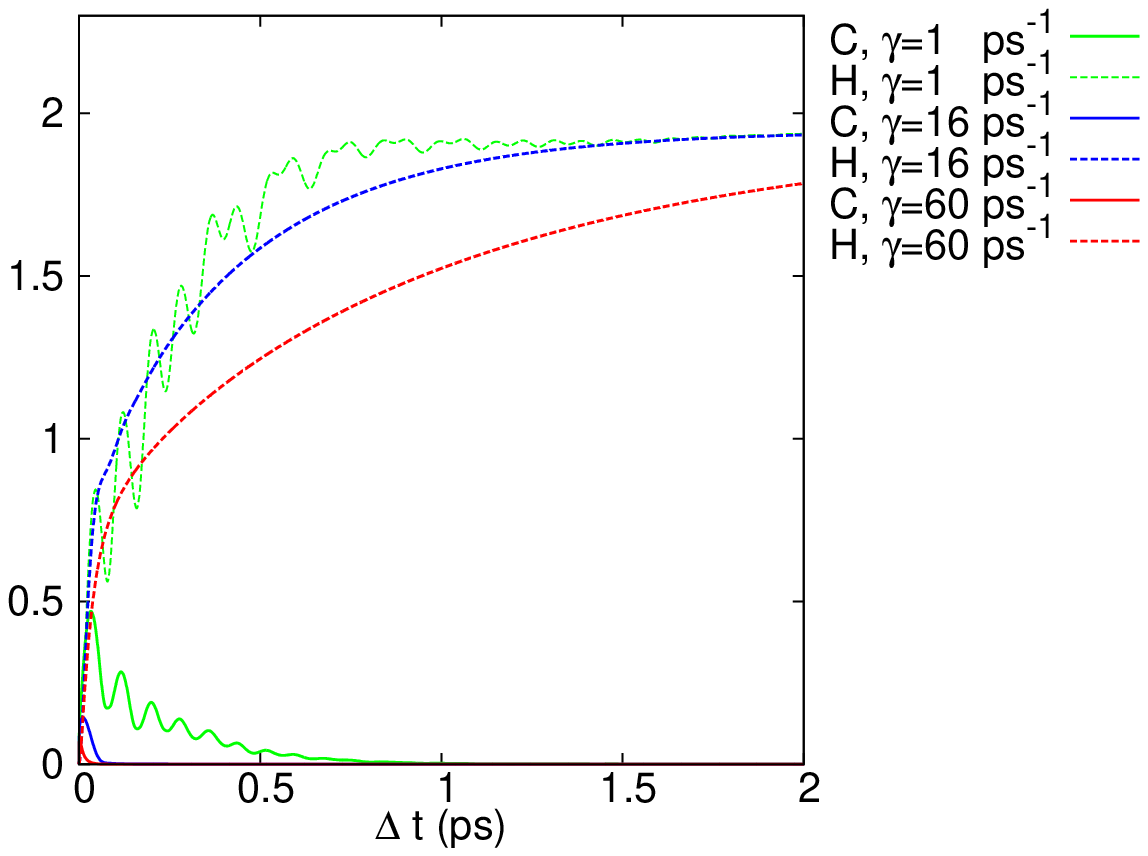}}
\protect\caption{FMO: \textbf{(a)} $\mathcal{C}(\Delta t)$ for different values
of $\gamma$; \textbf{(b)} $\mathcal{I}_{3}(\Delta t)$ for different
values of $\gamma$; \textbf{(c)} $\mathcal{H}(\tau=N\Delta t),\mathcal{C}(\Delta t)$
for $\gamma=1,16,60\:ps^{-1}$;}

\label{Fig: FMO C various gammas and C vs H} 
\end{figure}

\begin{figure}[t]
\subfigure[]{\includegraphics[width=0.4\textwidth]{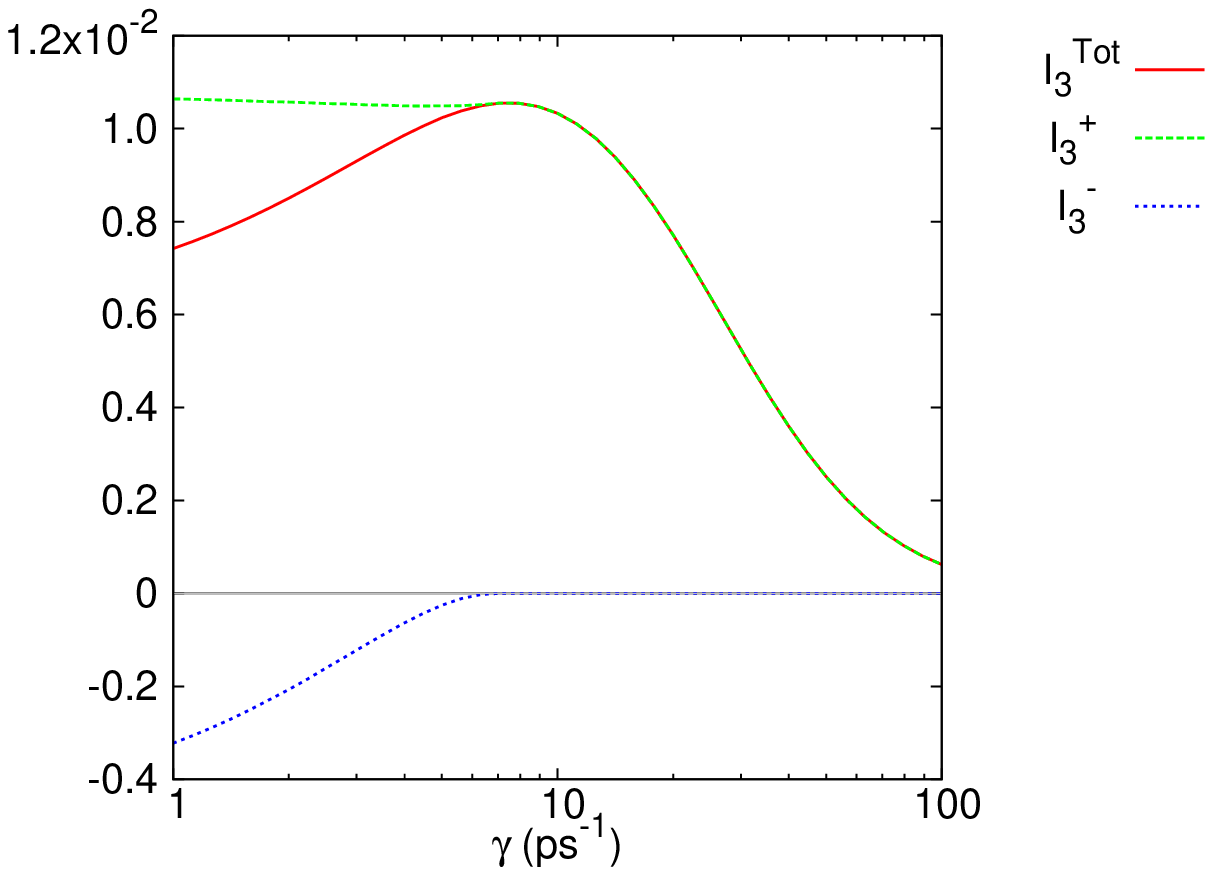}}
\subfigure[]{\includegraphics[width=0.4\textwidth]{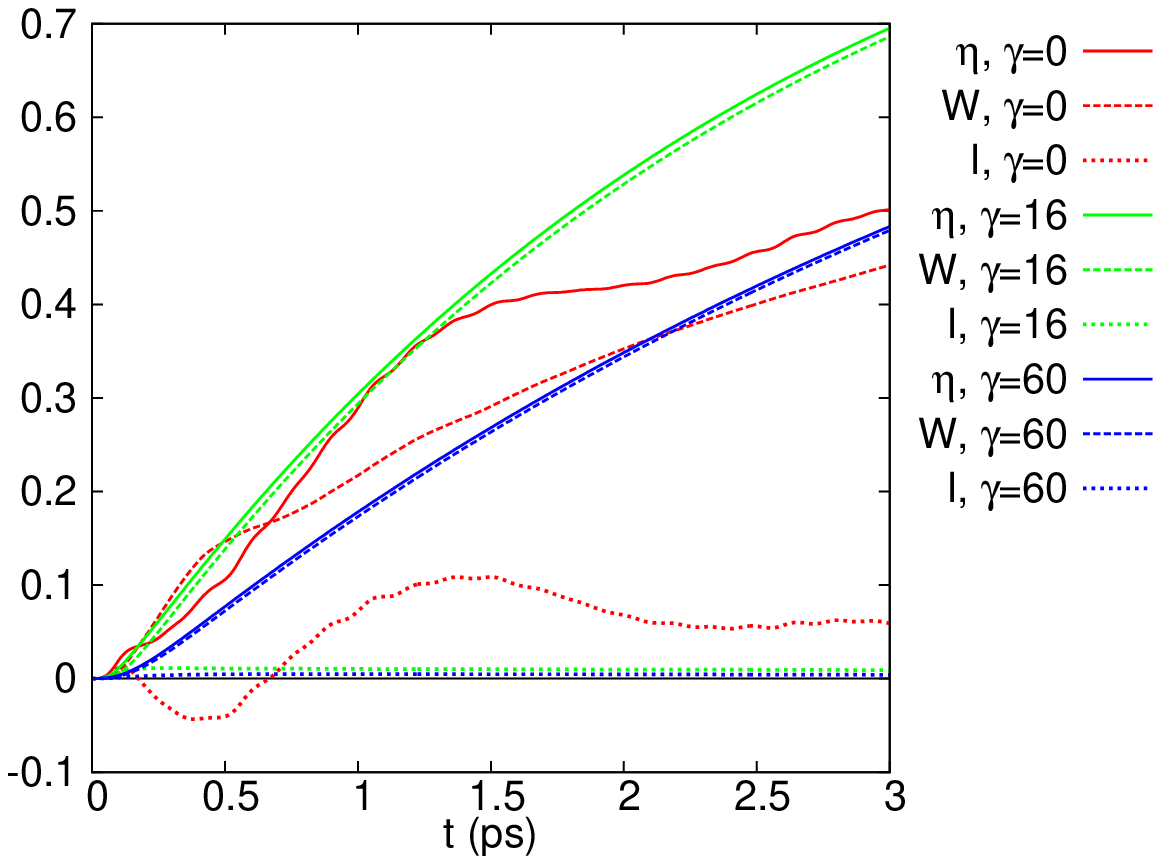}}
\protect\caption{FMO: \textbf{(a)} average interference $<\mathcal{I}_{3}>$ for
$k_{trap}=5\,ps^{-1}$; \textbf{(b)} transport efficiency $\eta(t)$, integrated weight $W_{3}(t)$ and
integrated interference $I_{3}(t)$ for pathways
ending at site $3$ for different values of $\gamma=0,1,16$ $ps^{-1}$ and
$k_{trap}=5\,ps^{-1}$}

\label{Fig: FMO average I on 200 fs 3 site and efficiency} 
\end{figure}

The above arguments can be easily applied to the whole FMO complex.
Fig.~\ref{Fig: FMO C various gammas and C vs H} and~\ref{Fig: FMO average I on 200 fs 3 site and efficiency}
show the application of the decoherent histories method to excitonic
transport in FMO. The main features of the behavior of $\mathcal{{C}},\mathcal{{H}},\mathcal{{Q}},\mathcal{{I}}_{3}$
and $\eta$ are maintained although obvious differences can be found
since the dynamics in now determined by the interplay of different
eigenperiods and interference paths are more complex. In particular,
Fig.\ref{Fig: FMO C various gammas and C vs H}(b), one can
observe a revival of positive interference $\mathcal{{I}}_{3}$ for
small values of $\gamma\simeq1$ , that does enhance the efficiency
for $t\approx1.5ps$, Fig. \ref{Fig: FMO average I on 200 fs 3 site and efficiency}(a); 
but this is not sufficient to compensate the initial and subsequent
negative interference, thus impeding the reach of optimal values of
$\eta$. In general, compared to the trimer and as suggested by Fig.
\ref{Fig: FMO average I on 200 fs 3 site and efficiency}(a) the
maximum average positive coherence on short time scales is attained
for smaller values of $\gamma$. The overall picture is not significantly
affected if one decides to start the dynamics from site $6$ instead
of site $1$, as it often is reported in the literature.

\section{Conclusions\label{sec:Conclusions}}

The decoherent histories approach provides a general theory to study
the distinctive feature exhibited by quantum systems: coherence. However,
despite its generality and foundational character, in order to measure
the effects of coherence and decoherence the DH approach needs to
be complemented with a quantitative way to condense the information
contained in the basic object of the theory, i.e., the decoherence
matrix $\mathcal{D}$. In this paper we introduce a set of tools that
allow one to assess the (global) coherence properties of quantum (Markovian)
evolution and that can be used to relate the coherence content of
a general quantum dynamical process to the relevant figure of merits
of the given problem. We first define the \textit{coherence functional}
$C(P,N,\Delta t)$, that can be interpreted as a measure of the global
coherence exhibited by the dynamics in the basis $P$ over the time
scale\emph{ $\Delta t$.} While this measure is completely general,
one can further introduce other relevant tools tailored to the specific
system and type of system-environment interaction at hand. We thus
focus on a simple yet paradigmatic model of environmentally assisted
energy transfer where coherence effects have been shown to play a
significant role in determining the efficiency of the process: a trimeric
subunit of the Fenna-Matthews-Olson photosynthetic complex. Based
on $\mathcal{D}$ and $C(P,N,\Delta t)$ we define: $a)$ a measure
$\mathcal{Q}_{\tau}(\gamma)$ able to characterize the average coherence
exhibited by the dynamics of the system over the time scales $\Delta t\in(0,\tau)$
for a fixed value of the dephasing $\gamma$; $b)$ a measure of the
average interference $<\mathcal{I}_{i}>$ occurring between the histories
ending at a given ``site'' $i$. \\Within the specific model, we
first thoroughly assess the consistency of the behavior of $C(P,N,\Delta t)$
in the various regimes. We then show how the introduced tools allow
to study the intricate connections between the efficiency of the transport
process and the coherence properties of the dynamics. In particular
we show that the delocalization of the exciton over the chromophoric
subunit is strongly affected by the amount of (average) coherence
allowed by the interaction with the bath in the first tens to hundreds
of femtoseconds. If the system-bath interaction is too strong, coherence
is suppressed alongside the interference between different histories,
in particular those ending at the site where the excitation leaves
the complex. If the interaction is too weak the system exhibits high
values of coherence even on long time scales, but it also exhibits
negative interference between pathways ending at the exit site, a
manifestation of the fact that the exciton bounces back and forth
over the network thus preventing its efficient extraction. In the
intermediate regime i.e., when the different time scales of the system
(quantum oscillations, decoherence and trapping rate) converge, the
system shows high values of coherence on those time scales. The action
of the bath has a \textit{quantum recoil avoiding effect} on the dynamics
of the excitation: the \textit{benefits of the fast initial quantum
delocalization of the exciton over the network are preserved and sustained
in time by the dynamics}; in terms of pathways leading to the exit
site, the action is to s\textit{electively kill the negative interference
between pathways, while retaining the initial positive one.} These
effects can be explicitly connected to the overall efficiency of the
environment-assisted quantum transport: the gain in efficiency
for intermediate (optimal) values of decoherence can thus be traced
back to the basic concepts of coherence and interference between pathways
as expressed in the decoherent histories language. \\While the specific
decoherence model used (Haken-Strobl) is an oversimplified description
of the actual dynamics taking place in real systems, we believe 
that our analysis allows to spot out the essential features that may
determine the high efficiency of the transport even in more complex
system-environment scenarios. \\In conclusion, the tools introduced
in this paper allow to thoroughly assess the coherence properties
of quantum evolutions and can be applied to a large variety of quantum
systems, the only limits being the restriction to Markovian dynamics
and the computational efforts required for high dimensional systems.
However, the extension to non-Markovian realms is indeed possible~\cite{Allegraprep},
and the use of parallel computing may allow the treatement of reasonably
large systems.

\begin{acknowledgements} 

P.~Giorda and M.~Allegra would like to
thank Dr.~Giorgio ``Giorgione'' Villosio for his friendship, his
support, his always reinvigorating optimism, and his warm hospitality
at the Institute for Women and Religion - Turin, where this paper
was completed (\textit{cogitato, mus pusillus quam sit sapiens bestia,
aetatem qui non cubili uni umquam committit suam, quin, si unum obsideatur,
aliud iam perfugium elegerit}).\\
 \ \\
P.~Giorda would like to thank Prof.~A.~Montorsi, Prof.~M.G.A.~Paris and  Prof.~M.~Genovese for their kind help. \\
 \ \\
S.~Lloyd would like to thank M.~Gell-Mann for helpful discussions.

\end{acknowledgements}

\end{document}